\documentclass[prd,twocolumn,aps,preprintnumbers,letterpaper,showpacs]{revtex4}
\usepackage{amsmath}
\usepackage{tipa}
\usepackage{amssymb}
\usepackage{graphicx}
\usepackage{bm}
\usepackage{enumerate}
\usepackage{color}
\usepackage[colorlinks = true,
            linkcolor = blue,
            urlcolor  = blue,
            citecolor = blue,
            anchorcolor = blue]{hyperref}
\usepackage{subfigure}

\newcommand{\be}{\begin{equation}}
\newcommand{\ee}{\end{equation}}
\newcommand{\bea}{\begin{eqnarray}}
\newcommand{\eea}{\end{eqnarray}}
\newcommand{\ba}{\begin{array}}
\newcommand{\ea}{\end{array}}
\newcommand{\beal}{\begin{align}}
\newcommand{\eeal}{\end{align}}

\newcommand{\nn}{\nonumber}

\newcommand{\ket}[1]{\left|#1\right\rangle}
\newcommand{\bra}[1]{\left\langle#1\right|}

\def\czn{c^{(\zeta)}_n }
\def\cxn{c^{(\xi)}_n }
\def\dzn{d^{(\zeta)}_n }
\def\h0{H^{(0)} }
\def\cz{c^{(\zeta)} }
\def\cx{c^{(\xi)} }
\def\dz{d^{(\zeta)} }
\def\hri#1#2{\href{http://arxiv.org/abs/#1}{[ArXiv:#1]#2}}
\def\hre#1#2{\href{http://arxiv.org/abs/#1/#2}{[ArXiv:#1/#2]}}

\begin{document}

\title{Collective String Interactions in AdS/QCD \\ and High Multiplicity $pA$ Collisions}
\author{Ioannis Iatrakis}
\email[Email: ]{ioannis.iatrakis@stonybrook.edu}
\author{Adith Ramamurti}
\email[Email: ]{adith.ramamurti@stonybrook.edu}
\author{Edward Shuryak}
\email[Email: ]{edward.shuryak@stonybrook.edu}
\affiliation{Department of Physics and Astronomy, Stony Brook University,\\ Stony Brook, New York 11794-3800, USA}

\date{\today}

\begin{abstract}
QCD strings originate from high-energy scattering in the form of Reggeons and Pomerons, and have been studied in some detail in lattice numerical simulations. Production of multiple strings, with their subsequent breaking, is now a mainstream model of high energy $pp$ and $pA$ collisions. Recent LHC experiments revealed that high multiplicity end of such collisions show interesting collective effects. This ignited an interest in the  interaction of QCD strings and multi-string dynamics. Holographic models, collectively known as AdS/QCD, developed in the last decade, describe both hadronic spectroscopy and basic thermodynamics, but so far no studies of the QCD strings have been done in this context. The subject of this paper is to do this. First, we study in more detail the scalar sector of hadronic spectroscopy, identifying ``glueballs" and ``scalar mesons," and calculate the degree of their mixing. The QCD strings, holographic images of the fundamental strings, thus have a ``gluonic core" and a ``sigma cloud." The latter generates $\sigma$ exchanges and collectivization of the strings, affecting, at a certain density, the chiral condensate and even the minimum of the effective string potential, responsible for the very existence of the QCD strings. Finally, we run dynamical simulations of the multi-string systems, in the ``spaghetti" setting approximating  central $pA$ collisions, and specify conditions for  their  collectivization into a black hole, or the dual QGP fireball.
\end{abstract}
\maketitle

\section{Introduction}

\subsection{QCD Strings and Holographic Models} 

High energy hadronic collisions in 1960's led to discoveries of Regge phenomenology of scattering amplitudes and striking Regge trajectories connecting mesonic and baryonic states. The leading singularity, producing a term in the amplitude \be A(s,t)\sim s^{\alpha(t)}\sim s^{\alpha(0)+\alpha'(0) t}\,, \ee is known as the $Pomeron$, named after I. Pomeranchuk. The Pomeron dominates the high-energy cross sections and elastic amplitudes. 
 
These observations were then explained in terms of the QCD strings. In particular, the slopes of the Regge trajectories, including that of the Pomeron, $\alpha'$, were related to the string tension. (Creators of the fundamental string theory have kept this notation for the fundamental string scale). String tension is also the slope of the linear potential of confinement, which has been studied in detail in lattice numerical simulations. Its value,  $T_s=(420\, \text{MeV})^2$, is in fact used as a standard definition of ``physical units" in various confining theories.

QCD strings have been with us for about half century, and yet, the interest to them was rather unsteady. From the lattice we learned that the QCD strings are surprisingly thin, with an r.m.s. width of only about $0.17$ fm. The dynamics of these strings are well described by the simplest Nambu-Goto action. The exponential growth of density of states of strings produces the Hagedorn phenomenon, a rapid excitation of the string, responsible for the deconfinement transition.

The discovery of QCD gave rise in 1970's to weak coupling, or pQCD, methods. In the field of hadronic collisions, the ``hard," or BFKL, Pomeron has been derived through the re-summation of gluonic  ladders. This approach focused on the Pomeron intercept $\alpha(0)$, while ignoring a ``stringy" $\alpha'(0)$. 

We return to the recent derivation of the Pomeron below, and now switch to strong coupling models, which came into existence after seminal discovery of the AdS/CFT duality. This relates the 4-dimensional,  $\mathcal{N} = 4$  super Yang-Mills theory at strong coupling and large number of colors with the weakly coupled Type IIB supergravity in an $AdS_5\times S^5$  background created by a set of $D_3$ branes, \cite{Maldacena:1997re}. 
 
 Maldacena in \cite{Maldacena:1997re} calculated the potential of two static charges connected by a string by deriving the shape of a bending (geodesic) string in the bulk spacetime. As the setting is scale invariant, the modified strong coupling Coulomb law remains $V\sim 1/r$, with only the coefficient modified.  The calculation for the charges moving with fixed velocity $\pm v$ away from each other has been performed by Lin and Shuryak \cite{Lin:2006rf}, who found that for small enough $v$ there is a scaling solution for the falling string, generalizing Maldacena case into a ``generalized Amp\`ere law." Yet, above certain critical velocity this becomes unstable, and the stable solution in that range has been found numerically. The hologram of that string calculated in \cite{Lin:2007pv} showed a near-spherical, non-hydrodynamical explosion, demonstrating quite directly that in $\cal{ N}$=4  at strong coupling there should be no jets, even in $e^+e^-$annihilation into quarks. Consequently, suppose one can describe a  set of falling strings in  $AdS_5$ (a ``strongly coupled glasma"). If the endpoints are separated by a very large rapidity interval (say $\sim \Delta y$=15 at the LHC), one would expect for the falling strings similarly large rapidities of the transverse flow. 

Both the gauge theory in question and the bulk $AdS_5$ spacetime possess conformal symmetry. While vital for the discovery of the duality, conformality  is certainly not welcomed when a description of the real world is intended. Over the years, the AdS/QCD approach has been developed, which includes violation of conformality.

%

Unlike lattice QCD, AdS/QCD is not restricted to the Euclidean domain, and thus various time-dependent  processes can also be studied. This opens a door to studies of various out-of-equilibrium settings devoted to understanding of the matter equilibration. For example, matter equilibration can be described via a ``shock wave in the 5$^\text{th}$ dimension", which is equivalent to a membrane falling under its own weight \cite{Lin:2008rw}.
 
So far, the AdS/QCD approach has been used to model the hadronic spectrum, as well as thermodynamics of QCD at finite temperatures. A particularly successful model of this class is the Improved Holographic QCD model, which incorporates several low energy QCD features like confinement and linearity of glueball trajectories, and reproduces the thermodynamic functions at finite temperature \cite{Gursoy:2007cb, Gursoy:2007er, Gursoy:2008za, Gursoy:2009jd}. The flavor degrees of freedom are usually studied in the probe approximation, where they do not backreact to the glue \cite{Erlich:2005qh, Erdmenger:2007cm,  ckp, ikp}. Important recent advances \cite{jk}, which are used below, include  the back-reaction of the quarks  to the glue in the Veneziano limit of QCD (also called V-QCD), in which the number of colors and flavors are comparable:
\be N_c,N_f \rightarrow \infty \, \,\,\, N_f/N_c=x=\text{const}  \, .\ee
Here, we are interested in the QCD-string interaction and the effect of the QCD strings on the chiral symmetry where $\sigma$-meson plays a crucial role since it mainly mediates the string interaction and it is responsible for chiral symmetry restoration, as explained in (\ref{scamaspheno}, \ref{chsymre}). Hence, it is necessary to work in the Veneziano limit where the flavor degrees of freedom are not suppressed. In the holographic context, the string directly sources the scalar glueballs, which are coupled to the $\sigma$-meson to leading order in the Veneziano limit \cite{Arean:2012mq, Arean:2013tja}.

\subsection{Strings in High Energy Collisions}
As has been already mentioned, high-energy collisions of hadrons are dominated by the Pomeron physics.
For recent derivation of the Pomeron amplitude in string-based holographic settings, see  \cite{Basar:2012jb,Shuryak:2013sra}. Without going into details, we remind the reader that the Pomeron process is basically a spontaneous {\em reconnection of the strings}; the hadrons, instead of being color neutral, are connected by (at least) two QCD strings after the collision. At the moment of the collisions, the strings are located in the transverse plane, appearing from a tunneling -- under the barrier -- process  \cite{Basar:2012jb}.  

High energy colliders --  RHIC at BNL and LHC at CERN -- continue to provide new data on a wide range of hadronic collisions, from the basic proton-proton ($pp$), to proton-nuclei ($dAu$ and $pAu$, to be called $pA$ below), and heavy-ion $AA$ collisions. As the multiplicity grows in each of the systems, a transition is observed. In the $AA$ case, peripheral collisions are superpositions of $NN$ collisions, while more central collisions lead to production of the QGP fireball, which subsequently explodes, leading to a set of observed collective phenomena, such as radial, elliptic, etc. flows. The typical $pp,\,pA$ events are, on the contrary, well explained by pQCD and QCD strings, decaying independently, according to Lund-type models, such as Pythia and its descendants. 

The question where the transition between those two regimes happens is currently under intensive study, and collective phenomena such as radial, elliptic, and triangular flows were indeed observed in high multiplicity $pA$, and perhaps even $pp$ collisions. 


For recent review of this phenomenology, one can see e.g. \cite{Shuryak:2014zxa}. Let us mention the main details of $pA$ collisions. Note that  the typical impact parameter,
\be b\sim \sqrt{\sigma_{pp}(s) \over \pi} \approx 1.6\,\text{fm}\,, \ee
(in the LHC energy range) is about one order of magnitude larger than the string width of $.17$ fm extracted from the lattice data. The diluteness of the QCD strings in such a state is given by the ratio 
\be  d=N_{s} \left({ r_\text{string}\over b}\right)^2  \sim 10^{-2} N_s\,. \ee
So, we see that even for $N_s\sim 40$ in the central $pA$ collision at the LHC, $d<1$ and the strings are not yet fully overlapping.

The central issue -- referred to as the ``radial flow puzzle" -- explains why we are so interested in $pA$ collisions. The initial density (multiplicity per transverse area $dN/dA_\perp$)  suggests the following order of magnitude of various processes,
\be {dN^{pA}_{central} \over dA_\perp} \sim {dN^{AA}_{peripheral} \over dA_\perp}  < {dN^{AA}_{central} \over dA_\perp} <  {dN^{pp}_{highest} \over dA_\perp}\,,  \ee   
and, na\"{i}vely, one expects that the magnitude of the explosion (the velocity or rapidity $v_\perp=\tanh(y_\perp)$ of the radial flow)  would also follow this pattern. And yet, experiment gives us a different trend,
\be y_\perp^{AA,peripheral} <  y_\perp^{AA,central} <  y_\perp^{pA,central} \sim y_\perp^{pp,highest}\,. \ee  
Central $pA$ collisions are therefore more ``explosive" than expected.

One scenario proposed to explain this puzzle, proposed by Kalaydzhyan and Shuryak \cite{Kalaydzhyan:2014zqa}, is a collective collapse of the string system created in maximal multiplicity $pA$ case. If so, the size of the system is reduced and density is increased by a significant factor, leading to a different order of the densities, 
corresponding to the strength of the flow. Another notable consequence of this collapse is that the combined 
field of all of the strings becomes strong enough to restore chiral symmetry and thus create a QGP fireball, needed for an explosion.

As time goes on and the remnants of the nucleons move forward and backward with large rapidities, the ends of the strings remain attached to them. As a result, the strings get extended longitudinally, and eventually break via quark pair production. The details of this process are modeled by descendants of the Lund model of particle production, such as the Pythia event generator. Since quark pair production proceeds via tunneling in the Schwinger mechanism, it is numerically suppressed. This allows a certain proper time $\tau_{breaking}\sim 1.5$ fm for strings to be stretched along the beam axis, forming a multi-string state. 
 
Since the creation of QGP fireball is dual to formation of a black hole in a bulk, this topic has been addressed in some of these works. We, however, focus instead on smaller systems such as typical $pp$ or $pA$ collisions, and thus instead of colliding walls, we discuss a string setting. Looking for a more amenable geometry, we look at the ``spaghetti" phase, in which certain number of $parallel$ QCD strings are produced and subsequently decay. The smallest number, originating from a single Pomeron or color exchange, is 2 strings, connected to leading quarks (diquarks).  

It has been argued in \cite{Kalaydzhyan:2014zqa} that when the number of strings in the ``spaghetti" gets large enough, mutual attraction between them grows enough to induce a collapse and the collectivization of their fields. It has been argued that it should happen for the largest number in observed high multiplicity $pA$ events, in which perhaps 40 or so strings are produced. In this paper, we extend this scenario into a holographic setting, where we calculate the string potential from the underlying holographic model.

\section{The Setting}
\subsection{Background Solutions} 
\label{backsol}
The IHQCD model was developed by Kiritsis et al. in  \cite{Gursoy:2007cb, Gursoy:2007er, Gursoy:2008za, Gursoy:2009jd}. This is a phenomenological holographic model of 4-dimensional Yang Mills at large $N_c$. It includes an asymptotically $AdS_5$ metric with logarithmic corrections corresponding to the perturbative running of the 't Hooft coupling at high energies. The model also contains a scalar dilaton field, which is the holographic analogue of the 't Hooft coupling. A sophisticated choice of the dilaton potential leads to the correct description of certain low-energy features of the theory. The mesonic sector of QCD is modeled by placing $N_f$ coincident pairs of $D4$-$\bar{D}4$ branes in the background metric. As was shown in \cite{ckp} and further analyzed in \cite{ikp}, chiral symmetry is broken as soon as the background metric is confining.  

The IHQCD model was generalized in \cite{jk} by including the back-reaction of the quarks, in a model named V-QCD. Its thermodynamics were studied in \cite{Alho:2012mh} and different aspects of its spectra in \cite{Arean:2012mq, Arean:2013tja, Iatrakis:2014txa, Jarvinen:2015ofa}. For convenience of the reader, the definitions and terminology of the fields involved are explained in Table \ref{bkgandflds}.

\begin{figure}[!tb]
\begin{center}
\includegraphics[width=0.49\textwidth]{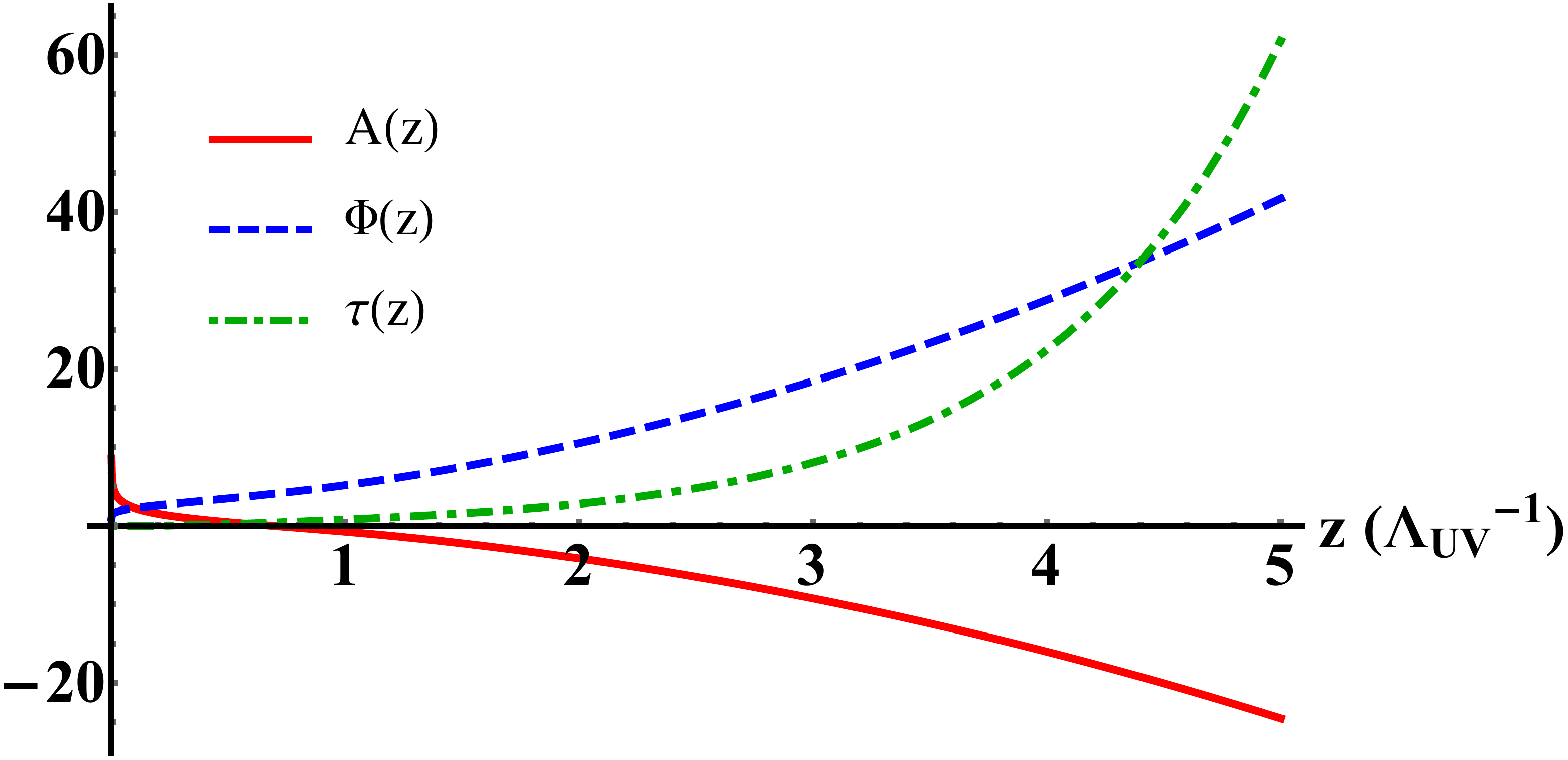}
\end{center}
\caption{The background solutions for the metric scale factor (solid), the dilaton (dashed) and the tachyon (dotted) as a function of the holographic coordinate $z$.}
\label{backsols}
\end{figure}

The action for the metric and the dilaton $\Phi$ is
\be S_g=M^3 N_c^2 \int d^5x \sqrt{-g} [R-{4 \over 3} g^{\mu\nu}\partial_\mu \Phi \partial_\nu \Phi +V(\Phi) ] \,,
\ee 
where $M$ is the 5-dimensional Planck mass.
The overall setting includes a background with a metric of the form 
\be 
g_{\mu\nu}=\exp(2A(z)) [dz^2+ \eta_{ij} dx^i dx^j ]  \,,
\label{conf_factor}
\ee
where $\eta_{ij}=\text{diag}(-,+,+,+)$ is the Minkowski metric. The 't Hooft $\lambda$ coupling is directly related to the dilaton: $\lambda=\exp(\Phi) $. The main field for the description of the flavor part of the theory is the tachyon ($\tau$), which is a bifundamental scalar under the $SU(N_f)_L\times SU(N_f)_R$ flavor symmetry and is dual to $\bar{q} q$ operator. The background fields are shown in Fig. \ref{backsols}. 

The action describing  the vacuum of the flavor sector has been analyzed in \cite{jk} and is of the form
\be
S_f=-M^3 N_c N_f \int d^5x V_f(\lambda, \tau)\sqrt{\det(g_{\mu\nu} + \kappa(\lambda) \partial_\mu \tau \partial_\nu \tau)}\, , 
\ee
where $V_f(\lambda,\tau)$ and $\kappa (\lambda)$ are potentials which were constrained by matching to QCD features in  \cite{Arean:2013tja}. Here we use the class of potentials I, which are defined in Eqs. (4.19) and (4.20) of   \cite{Arean:2013tja}, with $V_f(0,0)=3/11$.

The UV expansions of the background fields are such that they reproduce the perturbative running of the 't Hooft coupling. 
\begin{align}
A(z) & \sim -\log\frac{z}{\ell} + \frac{4}{9 \log(z \Lambda_{UV})} \,,\,\,  \lambda(z)\sim-\frac{8}{9V_1\, \log(z \Lambda_{UV})} \, , \nonumber \\
   \tau(z) &\sim \ell \, m_q  \,z \, (-\log(\Lambda_{UV}\,z))^{\gamma} + \ell \, \sigma_q \,z^3(-\log(\Lambda_{UV}\,z))^{-\gamma} \, ,
\label{tachuv}
\end{align}
where $\ell$ is the AdS radius and $\Lambda_{UV}$ is the characteristic scale of the theory that corresponds to the QCD strong coupling scale. The parameter $\gamma$ is determined by the anomalous dimension of $\bar q q$, \cite{jk}, $m_q$ is the bare quark mass, and $\sigma_q$ is the condensate. In the present work, we consider $m_q=0$. The determination of $\sigma_q$ is numerically very challenging; see \cite{Jarvinen:2015ofa}. The IR asymptotics of the fields are

\begin{align}
\label{irback}
 A(z) \sim -\Lambda_{IR}^2 z^2+& {1\over2 }\log(\Lambda_{IR} z) \,\, , \,\,\, \Phi(z) \sim {3\over 2} \Lambda_{IR}^2 z^2 \, ,\\
 & \tau(z) \sim \tau_0 e^{C \, \Lambda_{IR} z }\, ,
\end{align}
where the IR scale $\Lambda_{IR}$ is an integration constant of the equations of motion.

The scalar fluctuations of the model describe the mixed scalar glueballs and mesons and were analyzed in \cite{Arean:2013tja}. The relevant excitations of the tachyon, dilaton, and the scalar part of the the $g_{ij}$ component of the metric are
\begin{align}
\Phi=\Phi+\chi\,\,, \,\,\, T=\tau + s \,\,,\,\,\, g_{ij}=e^{2 A} (1+2 \psi) \eta_{ij} \,.
\end{align}
Their invariant combinations under the linearized 5-dimensional diffeomorphism symmetry of the bulk space-time which correspond to the physical scalars are 
\be
\zeta=\psi-{A' \over \Phi'} \chi \,\,, \,\,\, \xi=\psi-{A' \over \tau'} s \, .
\ee
$\zeta$ and $\xi $ are dual to the RG operators of the boundary field theory which generate the glueball and scalar mesons, respectively, as $x \to 0$.  In the Veneziano limit $(x=\rm{finite})$, the two excitations mix to leading order, hence the distinction to glueballs and mesons requires a detailed study of the mixed fluctuation equations, (\ref{zeteq}, \ref{xieq}). We perform this analysis in section (\ref{mixsec}) for $x=1$ and make contact with phenomenology. We show that the mixing of the different fluctuations can be weak even for finite value of $x$.

\begin{table}[htdp]
\caption{The fields and fluctuations of our model.}
\begin{center}
\begin{tabular}{|c|c|l|}
\hline
 $\Phi= \log \lambda $ & Dilaton \\
$A$ &  Metric conformal factor  \\
$A_s =A+{2 \over 3} \Phi$ & String frame metric  conformal factor\\
$\tau$ &  Tachyon ($\bar{q}q$ scalar)  \\ 
\hline
$\chi=\delta \Phi$   & Dilaton fluctuation \\
$s =\delta \tau$ &  Tachyon fluctuation \\
$\psi = {e^{-2 A} \over 4} \delta g_i^i $ & Scalar part of metric fluctuation \\
\hline
$\zeta = \psi-{A' \over \Phi'} \chi $  & Scalar glueballs as $x \to 0$\\
$\xi = \psi-{A' \over \tau'} s $ & Scalar  mesons as $x\to 0$\\

\hline
\end{tabular}
\end{center}
\label{bkgandflds}
\end{table}%

%
%
%

\subsection{Strings}
The bulk theory is argued to be a low energy effective model of string theory, even if we are not in a control string theory limit, so there are fundamental strings in the bulk. The electric fluxes running through these strings are sourced by charges at the boundary -- the quarks of the gauge theory. In holographic models, studies of stretching strings were first done in the original AdS/CFT correspondence with a conformal $AdS_5$ space in refs. \cite{Lin:2006rf,Lin:2007fa}. In this setting, the middle section of the string is falling under gravity away from the boundary. The calculated holographic image of such stretching strings reveals a non-hydrodynamical explosion, which does not have any ``jet-like" features (forward-backward peaks).

The bulk strings in AdS/QCD background are not indefinitely falling, but can reach certain equilibrium positions and ``levitate". Their hologram at the space boundary is what one would call the QCD strings. Thus in such models the forces between quarks gradually change from Coulomb $1/r$ at small $r$ to linear $V\sim r$ at large $r$, corresponding to confinement, \cite{Gursoy:2007er}. 

String dynamics are governed by the Nambu-Goto action 
\be S_{NG}=-T_f \int d\tau d\sigma \sqrt{- \det g_S} \, , \ee 
\be (g_S)_{\alpha\beta}=(g_S)_{\mu\nu} \partial_\alpha X^\mu \partial_\beta X^\nu \, , \ee 
where we use the string frame metric 
\be 
(g_S)_{\mu\nu}=e^{2A_s(z)}\eta_{\mu\nu}\,\,,\,\,\, A_s(z)=A(z)+{2\over 3} \Phi(z)  \, .
\ee

In top-down holographic models, the fundamental string tension $T_f$ provides the input scale for the whole construction and is determined by the original string theory. Unfortunately, in the V-QCD model, such a connection of $T_f$ to other parameters is missing, and thus should be fitted to phenomenology. 

\begin{figure}[!tb]
\begin{center}
\includegraphics[width=0.49\textwidth]{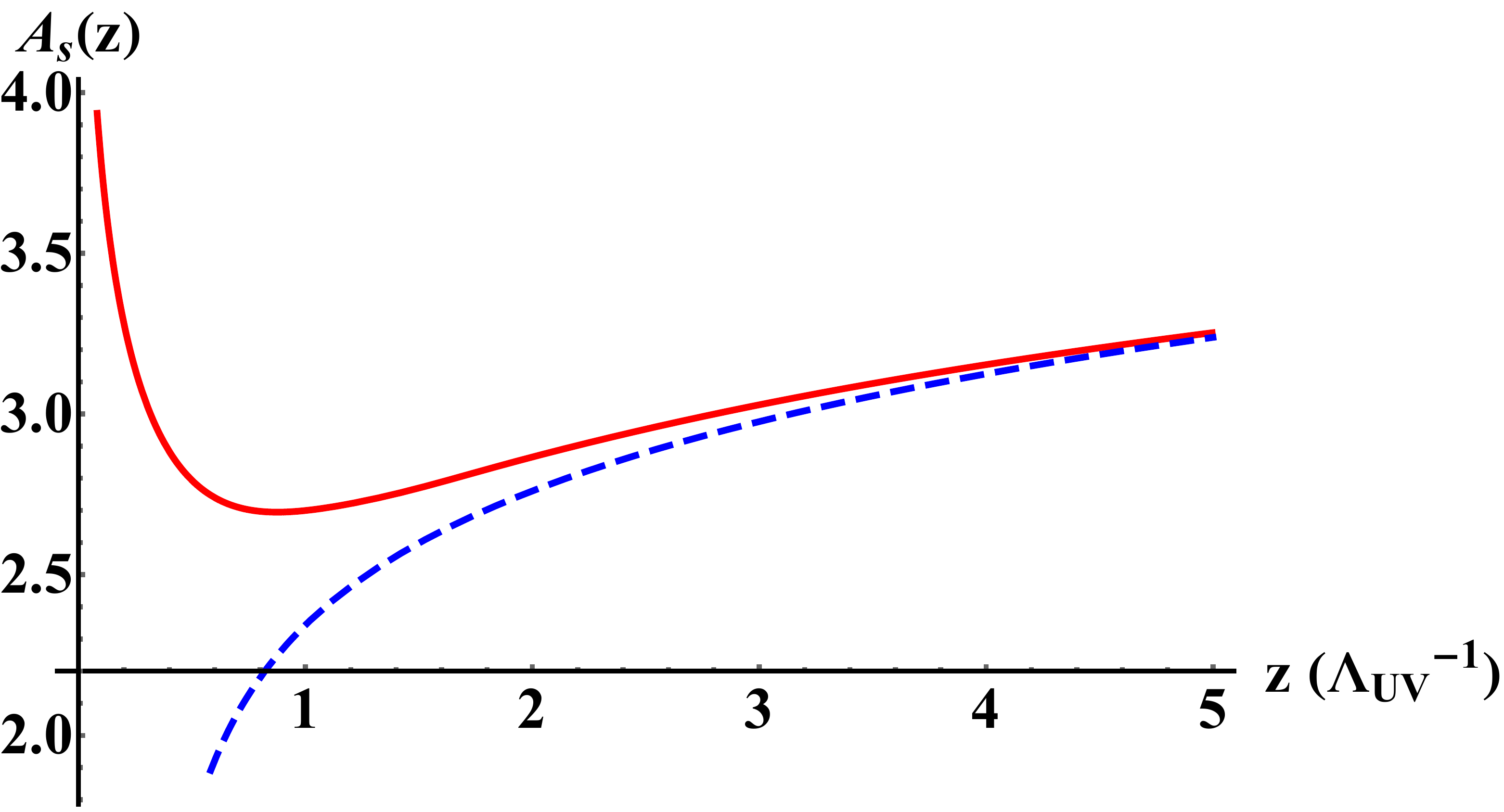}
\end{center}
\caption{The combination $A_s(z)$ as a function of the holographic coordinate $z$ (solid) compared to its IR (large $z$) asymptotics (dashed). $A_s(z)$ has a minimum corresponding to the equilibrium scale of the QCD string.  
}
\label{Asfig}
\end{figure}

The force of gravity -- gradients of the metric -- causes all objects, including strings, to fall to the IR (large $z$). The dilaton gradient, however, produces the opposite effect. Specifically, in the background used, the metric at large $z$  is decreasing as $A\sim - \Lambda_{IR} z^2$, but the $\Phi$ contribution cancels this term; see Eq. (\ref{irback}). The conformal factor of the string frame metric then increases in the IR, $A_s \sim (1/2)\ln(z)$ at large $z$. As a result, at some position $z_*$ there is a minimum of $A_s$ where the string equilibrates; see Fig. \ref{Asfig}. This minimum is given by
\be A_s'(z_*)=0,  \,\,z_*\approx 0.876 \, \Lambda_{UV}^{-1} \ee

The simplest falling string example is a string extended in $x$ direction and falling in $z$: $X^\mu=(t,x,0,0,Z(t)) $. Plugging this into the action, one finds the effective action of the problem for $Z(t)$: 
\be \int dt dx e^{2A_s(Z)} \sqrt{1- \dot{Z}^2} \, , \ee 
which generates the equation of motion (EOM): 
\be \partial_t \left( e^{2A_s(Z) }{ \dot{Z} \over  \sqrt{1- \dot{Z}^2} } \right) = \sqrt{1- \dot{Z}^2}   \partial_z \left( e^{2A_s(Z)} \right)  \,. \ee
Instead of solving this EOM, for one string one can use instead conservation of energy in our time-independent background. The Hamiltonian for this case is
\be H= -{\exp(2A_s(Z)) \over \sqrt{1-\dot Z^2} }  \,. \ee
By setting it equal to the energy $E$, one gets directly the first derivative of $Z$.

As usual, motion in general occurs between two turning points in which the velocity vanishes $\dot Z=0$.
When $z=z_*$ and the energy corresponds to the minimum, the string simply ``levitates" without motion at this point. The holographic image of this stationary string, calculated by standard rules, generates some stress tensor $T^{\mu\nu}(x)$ distribution, with Minkowski indices $\mu,\nu=0..3$, describing a static QCD string. The integral $\int d^3 x T^{00}(x)$ per unit length in $x$ is known as the QCD string tension $T_s=(420\, MeV)^2$. Rather than predicting it, one can use it to fix the fundamental string tension $T_f$, \cite{Gursoy:2009jd}. 

Near the levitation point one can approximate the motion by a harmonic oscillation with the frequency 
 \be \omega_*=[2  A_s''(z_*)]^{1/2} \,. \label{eqn_freq} \ee
However, because the potential is quite asymmetric, the validity of such harmonic approximation is rather limited. In general, oscillations toward large $z$ -- the IR -- take longer time and reach higher values of $z$, as compared to motion in the UV direction.

Let us now return to the discussion of the holographic image of the string on the boundary. Standard rules require calculation of the bulk-to-boundary propagators for all bulk fields, which source the operators of the gauge theory. One of them -- the metric $g_{\mu\nu}$ -- sources the stress tensor $T^{\mu\nu}$ mentioned above. The characteristic scale of the string energy distribution in transverse plane is given by the mass of the corresponding mode, known as the tensor glueball  $M(2^{++})\approx 2$ GeV. This (rather large) scale defines the scale of the QCD string width $\delta y \sim 1/M(2^{++})\approx 0.1$ fm.  

Flavorless scalars source the quark bilinear operator $\bar q q$, related to modification of the quark condensate in the QCD vacuum by the string. The lowest scalar meson in QCD is much lighter than the tensor glueball  $M(0^{++})=m_\sigma \approx 0.4$ GeV, and thus it produce much wider image.  This field will play prominent role in what follows.


The main subject of the present paper is not the motion of an individual string, but rather the interaction and collectivization of many. A convenient experimental setting is provided by $pA$ collisions. In peripheral collisions, an incoming proton collides with one or a few nucleons, producing several strings. But in central collisions the number of interacting, or ``wounded", nucleons is rather large, $N_w\sim 20$, producing at least $N_s\sim 40$ strings. Transition from a dilute to dense regime can thus be experimentally studied as a function of the multiplicity of the produced secondaries.

Furthermore, in real heavy-ion collisions ($pPb$ at LHC, $dAu,He^3 Au$ at RHIC), the multiplicity of secondaries depends linearly on the rapidity, with the ratio between the positive and negative 
rapidity ends being of the order 2. The Lund string model explains this by the fact that only a fraction of the strings go from valence quarks to valence quarks in the fragmentation regions of light and heavy nuclei; all others end at the gluons located at intermediate rapidities. 

For geometric simplicity, however, we will ignore such complications, and consider ensemble of strings which are parallel and infinitely long, so that their motion can be reduced to 3-dimensional coordinate normal to the time and beam directions, referred to as the ``spaghetti'' state. The number of strings we use should be understood as a local number at mid-rapidity, which is where most of the collider detectors and measurements are done. Such approximations we have inherited from recent study  \cite{Kalaydzhyan:2014zqa}, which first addressed the issue of string collectivization not in the context of AdS/QCD.


 \section{The Scalar Mesons}
 \subsection{The Scalar Masses in the Model vs. Phenomenology}
 \label{scamaspheno}
Before we turn to our model calculations, let us for completeness present a brief summary of the relevant hadronic spectroscopy. The scalar flavor-singlet sector of hadronic spectroscopy we need to address is one of the most complicated ones; its detailed discussion can be found in publications of Particle Data Group and the vast literature on which it is based \cite{Agashe:2014kda}.
  
The scalar meson channel includes one light state, known as the $\sigma$ meson, or $f_0(600)$ in the current PDG naming scheme. Multiple complex fits put its location around $M+i \Gamma/2 \approx (400-600) +i (200-300)$ MeV, but with a large spread. It is followed by a group of relatively close states $f_0(1370), f_0(1500)$, and $f_0(1710)$. In general, all of them are expected to be a mixture of $\bar u u+\bar d d,\,\bar s s$, and the lowest glueball states. A number of arguments -- narrow width, diffractive production, and absence in $\gamma\gamma$ production channel -- indicate that the middle state $f_0(1500)$ is mostly comprised of the glueball. Splitting of  $f_0(1370)$ and $f_0(1710)$ is comparable to $2m_s$, and their decay channel suggests that the upper state is predominantly a strange $\bar s s$ meson.
  
 
The chiral symmetry breaking phenomenon is strongly related with the existence of the light scalar meson $\sigma-f_0(600)$. The standard form of its interaction with quarks vacuum, $\sigma \bar q q$, shows that the vacuum expectation value (VEV) $\left<\sigma\right>$ is -- up to a sign -- nothing else but the ``constituent quark mass." Partial or full cancellation of this VEV implies local restoration of chiral symmetry.
 
 
It is crucial to our project to include the back-reactions of the quarks on the gluonic observables, or use the Veneziano limit. The study of the spectrum of such models has been recently done in  \cite{Arean:2012mq, Arean:2013tja, Iatrakis:2014txa, Jarvinen:2015ofa}. In particular, we use the confined phase spectrum with zero quark mass and the holographic potential of class I, which was analyzed in \cite{Arean:2013tja}. To calculate the meson and glueball spectra in holography, we have to expand the action to quadratic order and solve the linear Sturm-Liouville problem for the fluctuation fields in the bulk space-time. The numerical integration of the excitation equations starts from the IR with normalizable boundary conditions, and then proceeds towards the UV boundary of space, where we impose normalizable boundary conditions. Mass values which cause ``bad" (non-normalizable) solutions to vanish in the UV provide the spectrum of the model. The masses of the lowest five modes in the scalar channel can be read off of Fig.  \ref{scalarmasses}. The glueball spectrum has also been studied in the context of other holographic models, such as the Sakai-Sugimoto model, but in the limit of small number of flavors, \cite{Brunner:2015oqa}. 

As seen from this plot, and previously shown in Fig. 7 (left) of  \cite{Arean:2013tja}, at physically relevant flavor parameter $x=N_f/N_c\approx 1$, the lowest non-flavored scalar is indeed significantly lighter than the others. Fig. 8 (left) of \cite{Arean:2013tja} puts its mass to about half of the $\rho$ meson mass and in the phenomenologically expected $\sim 400$ MeV range.

Next come a close pair of the second and third states, with mass ratios to the first one $m_3/m_1\approx 2.6$. Since in the calculation the strange quark is as light as $u,\,d$, there should not be a separate $f_0(1710)$ state, and this pair can be identified with a close pair $f_0(1370), f_0(1500)$; at $x=N_f/N_c=1$ their splitting is also correct. Different $x$-dependence of the third state from others hints that it is indeed mostly a glueball, but this feature is not robust, as it depends on the details of the potential.
 
 The five lowest masses in units of the UV scale of the model are  
 \bea
&&{  m_1 \over \Lambda_{UV}}=1.53\,, \,\,\, {m_2 \over \Lambda_{UV}}= 3.54\,,\,\,\, {m_3 \over \Lambda_{UV}} =3.94\,,  \nonumber \\
&& {m_4 \over \Lambda_{UV}}=4.86 \,,\,\,\, {m_5 \over \Lambda_{UV}}=5.45 \, .
\eea

Selecting the numerical value of $\Lambda_{UV}$ from the second and third state masses, which are narrow and therefore well mapped to phenomenology, we fix the absolute units in our model to be  \be \Lambda_{UV}=387 \,\text{MeV}, \,\,\, m_\sigma= 592 \, \text{MeV.} \ee
Then, of the masses of the higher states read

\bea
&& m_2 = 1370 \, \text{MeV}\,,\,\,\, m_3 =1525 \, \text{MeV} \,,  \nonumber \\
&& m_4 =1881 \, \text{MeV} \,,\,\,\, m_5=2019 \, \text{MeV} \, .
\eea

\subsection{Mixing of Pure States}
\label{mixsec}

In the language of QCD, there are two kind of unflavored hadrons: those made of glue -- $glueballs$, and those made of quark-antiquark pairs -- $mesons$ ($\bar q q$). In the string language these are, respectively, made from closed and open strings. In the AdS/QCD models those originate from fluctuations of closed strings coming from the color branes or open strings from the flavor branes. 

In any of those pictures, states with identical quantum numbers can mix, and when the mixing is strong, the original designation of states loses its meaning and one can only follow evolution of those as  $x=N_f/N_c$ changes. However, if the mixing remains relatively weak, such a distinction of states remains meaningful and can help one to understand what is happening, in particular in the small-magnitude of the string-string interactions that we focus on in this work.
 
With such motivation, we have studied the mixing phenomenon in significant detail. The excitation equations were derived in \cite{Arean:2013tja},  Eqs. (A100, A101),  for the two gauge invariant combinations of scalars defined in (\ref{backsol}). We use them in  the following form, of two coupled equations
 
\begin{align}
\zeta''+ \tilde{k}(A) \,\zeta'+ \tilde{p}(A) \,\xi'+z'(A)^2 \, m^2  \zeta+\tilde{N_1}(A) \left(\zeta-\xi\right)&=0\,,\label{zeteq} \\
\xi''+\tilde{q}(A) \,\zeta'+ \tilde{n}(A)\,\xi'+\tilde{t}(A) \, m^2 \xi+\tilde{N_2}(A) \left(\xi-\zeta\right)&=0\,,\label{xieq} 
\end{align}
where the coordinate used is the scale factor $A$ instead of the usual AdS coordinate $z$ and the derivatives are with respect to $A$. This choice of coordinates is that the numerical calculations close to the AdS boundary become substantially more accurate. The coefficients in equations (\ref{zeteq}) and (\ref{xieq}) are, expressed in the $A$ coordinate:

\begin{align}
&\tilde{k}(A)=\left( k(z(A)) z'(A)-{z''(A) \over z'(A)^2} \right) \,,\,\, \tilde{p}(A)=p(z(A)) z'(A) \,,\\
&\tilde{n}(A)=\left( n(z(A)) z'(A)-{z''(A) \over z'(A)^2} \right) \, ,\,\, \tilde{q}= q(z(A)) z'(A) \, , \\ 
&\tilde{N}_1(A)=N_1(z(A)) z'(A)^2  , \, \tilde{N}_2(A)=N_2(z(A)) z'(A)^2 , \\
& \tilde t(A)=t(z(A))z'(A)^2 \, , 
\end{align}
where the original coefficients (without the tildes) are given by the lengthy expressions in \cite{Arean:2013tja}, Eqs. (A.102 - A.107). We will not copy them here, but just comment that all parameters of those coefficients are fixed by the Lagrangian, and so, even with all the complexity of expressions, everything is fully determined. Normalizable solutions, both in IR and UV, provide the scalar masses that we already discussed above. 

We defined a set of ``zeroth order" states $\zeta_n^{(0)},\xi_n^{(0)}$ as the eigenfunctions of the Hamiltonian without mixing
 \begin{equation}
H^{(0)} = \left( \begin{array}{cc}
H_\zeta & 0 \\
0 & H_\xi \\ \end{array} \right) \, .
\end{equation}
The eigenvalues are ``unmixed masses squared" $m^{(0) \, 2}_n$ and 
\begin{align}
& H_\zeta=-  \frac{w_\zeta (A)}{z'(A)^2} \left( \frac{d^2}{dA^2}+ \tilde{k}(A) \,\frac{d}{dA}  +\tilde{N_1}(A)  \right)  \, , \\
& H_\xi=-\frac{w_\xi (A)}{\tilde t(A)} \left(\frac{d^2}{dA^2}  + \tilde{n}(A)\, \frac{d}{dA}  +\tilde{N_2}(A) \right)   \, .
\end{align}
The normalization weight for each is found by the standard transformation to the Sturm-Liouville form of the uncoupled fluctuation equations, and all of those states are subsequently normalized to the unit norm. Hence, we have defined
\be
w_{\zeta} (A)=z'(A)^2 e^{\int \tilde{k}(A)} \, \, , \,\,\ w_{\xi} (A)=\tilde t(A)^2 e^{\int \tilde{n}(A)} \,.
\ee

\begin{figure}[!tb]
\begin{center}
\includegraphics[width=0.49\textwidth]{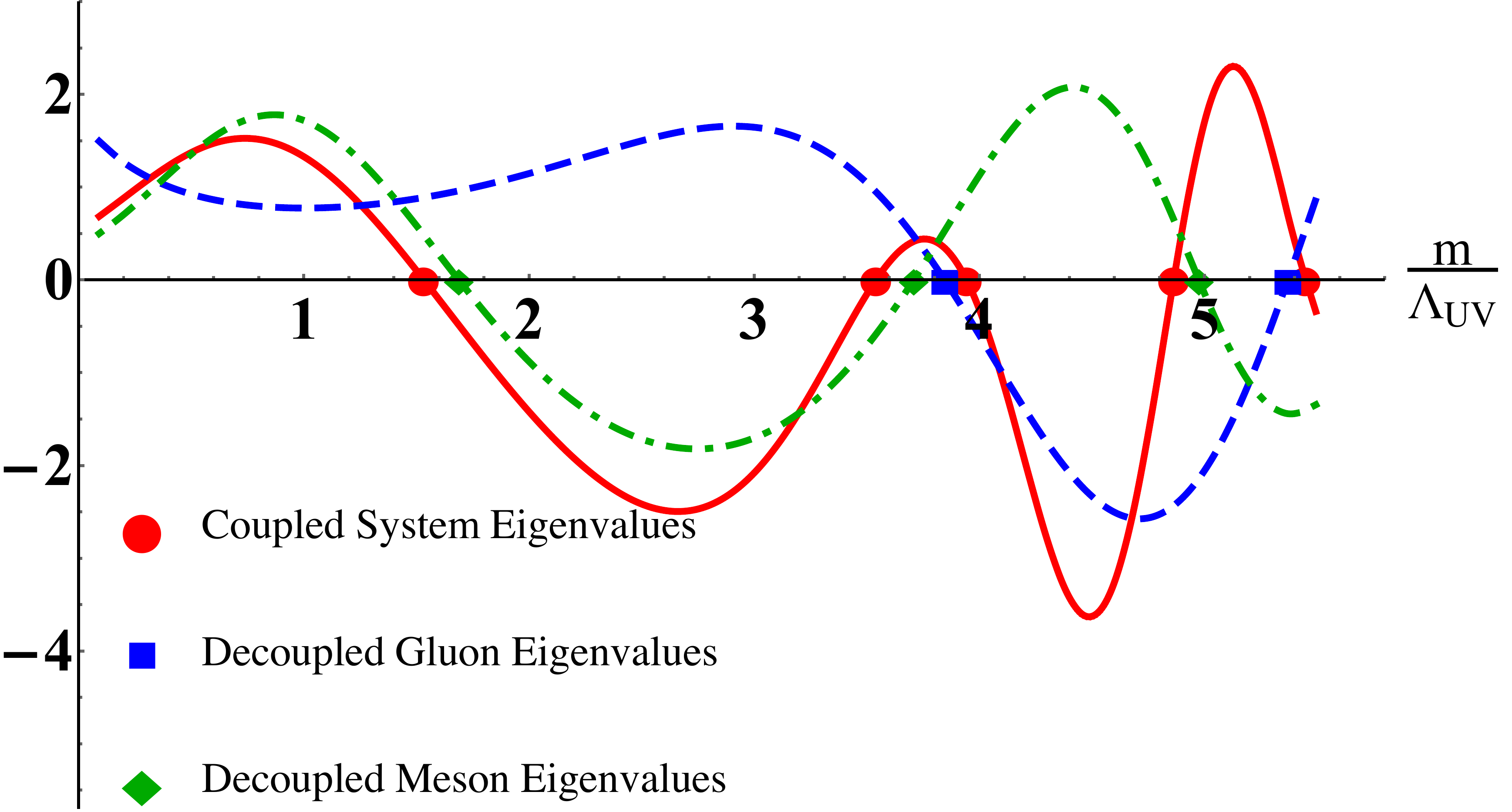}
\end{center}
\caption{The determinant of the UV boundary value of two linearly independent solutions of the scalar fluctuation equations, versus the mass parameter in  $\Lambda_{UV}$ units. The solid curve's five zeros (red circles) indicate the normalizable solutions, and the corresponding masses are those of the lowest five mixed scalars. The two other curves (dashed and dot-dashed) correspond to unmixed equations as explained in the text.}
\label{scalarmasses}
\end{figure}

In Fig. \ref{scalarmasses}, we show the determinant of the UV boundary value of two linearly independent solutions of the scalar fluctuation equations as a function of mass squared. Zeros correspond to the normalizable solutions and denote the eigenvalues. The curve with closed circles corresponds to fully-coupled system, giving the ``mixed mass squared," while two other curves are for uncoupled equations. The lesson from this plot is that each mixed state is close in its mass to one of the unmixed states we use as a basis; this was our early indication that mixing effects are, in some sense, small. Looking at the Fig. \ref{scalarmasses} more attentively, one finds an expected pattern of repulsive mass levels due to to mixing: close pairs of states move away from each other, the lowest state moves lower, etc.

\begin{figure}[!tb]
\begin{center}
\subfigure[]{%
\includegraphics[width=0.49\textwidth]{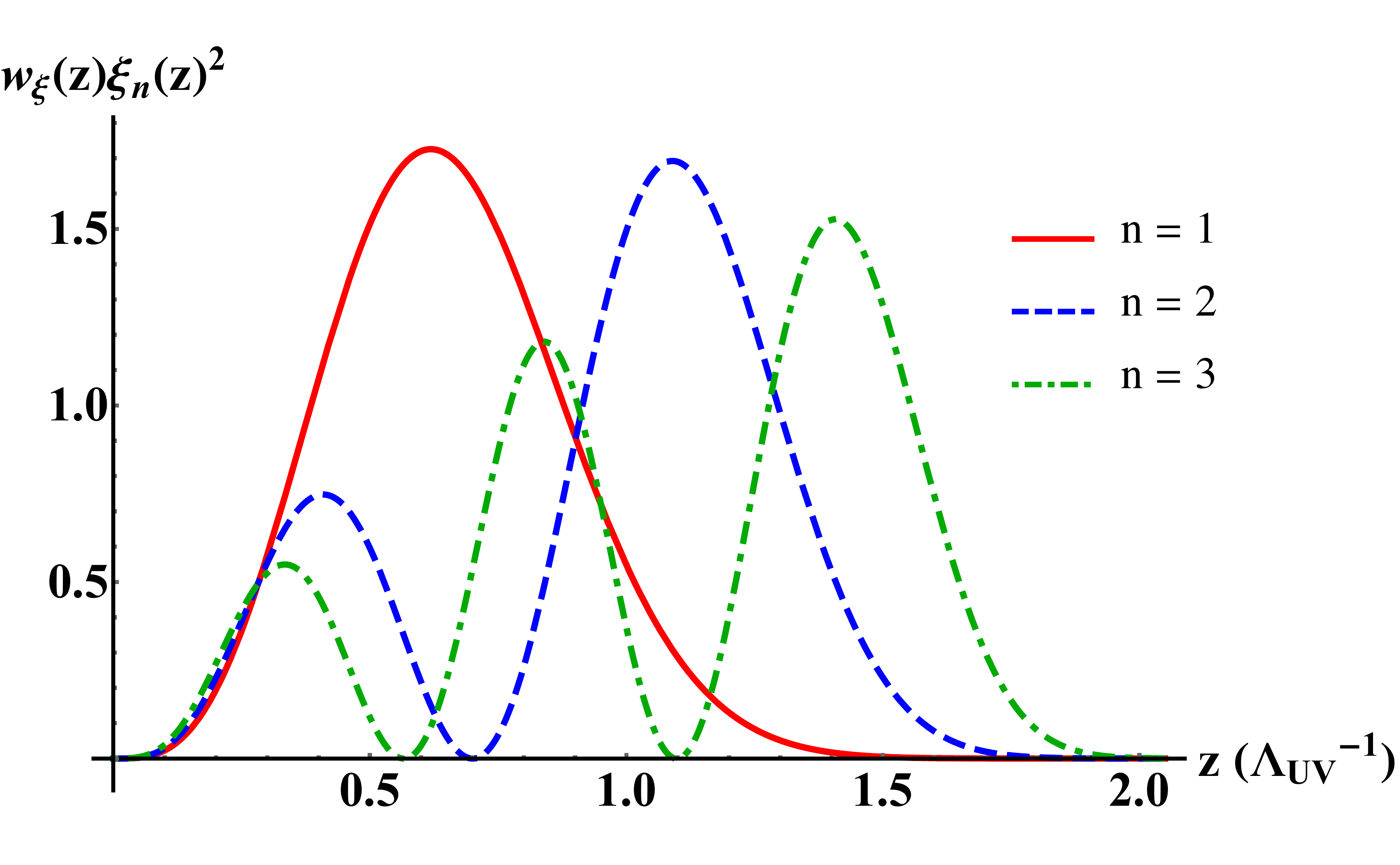}} \quad
\subfigure[]{
\includegraphics[width=0.49\textwidth]{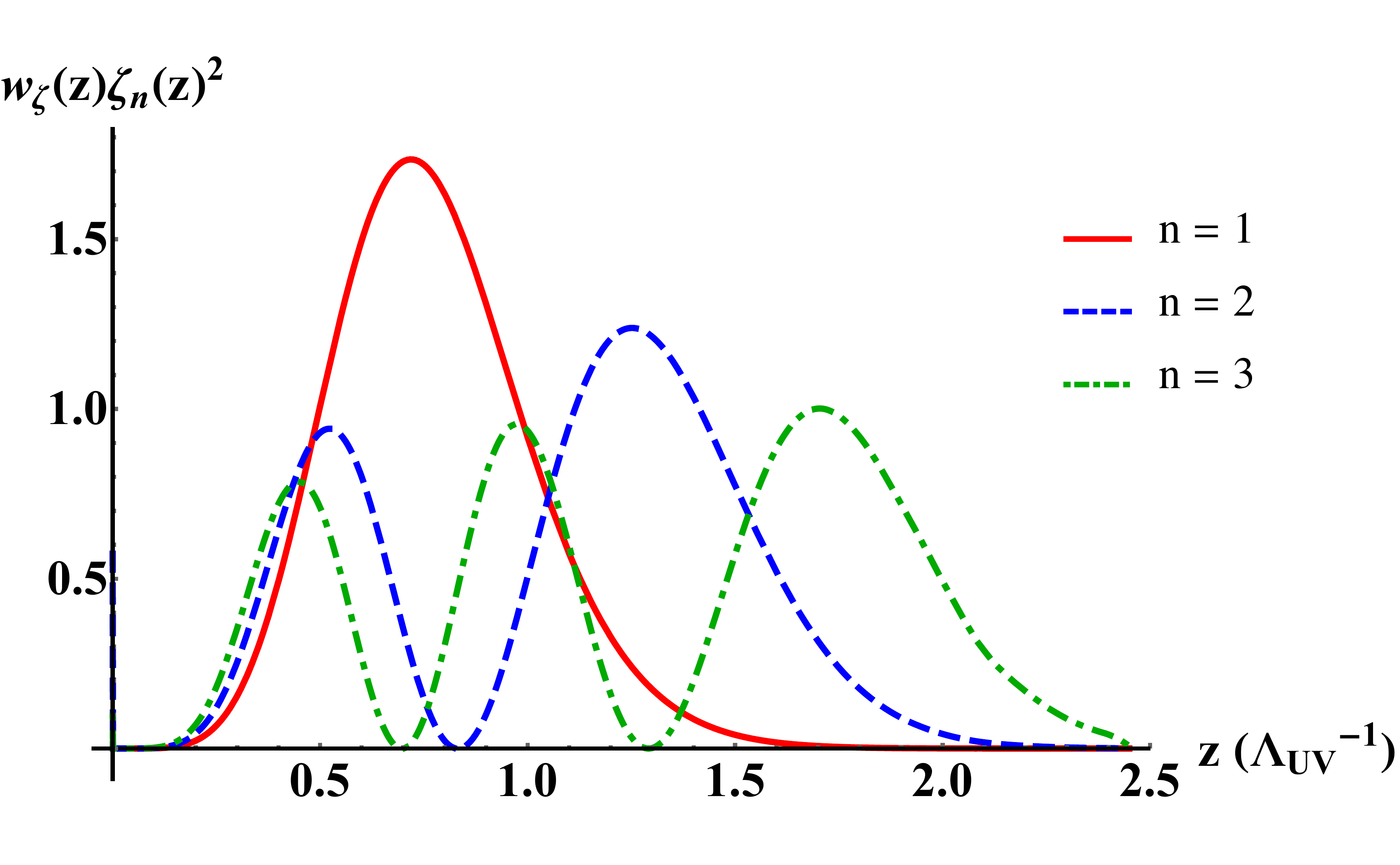}} 
\end{center}
\caption{The first three (a) meson and (b) glueball  wave functions squared, multiplied by the appropriate weight.}
\label{spect1}
\end{figure}

To make it quantitative we proceed by defining the mixing part of the Hamiltonian
\begin{equation}
V = \left( \begin{array}{cc}
0 & V_\xi  \\
V_\zeta & 0  \\ \end{array} \right) \, ,
\end{equation}
where 
\begin{align}
	V_{\zeta} &= -\frac{w_\zeta (A)}{z'(A)^2}\left(\tilde p(A) \frac{d}{dA} - \tilde N_1(A)\right)\,, \nn  \\
	V_\xi &= -\frac{w_\xi (A)}{\tilde{t}(A)^2}\left(\tilde q(A) \frac{d}{dA} - \tilde{N}_2(A)\right) \, .
\end{align}
Using unmixed states as a basis, we define a vector
\begin{equation*}
\Psi_{nm} = \left( \begin{array}{c}
( \zeta_n^{(0) })\\
(\xi_m^{(0)})\\
\end{array} \right) \, ,
\end{equation*}
where $m, n=1,2,3,\ldots$  We then calculate matrix elements $V_{nmn'm'}=\bra{\Psi_{nm}}V\ket{\Psi_{n'm'}}$ of the mixing. Limiting ourselves to the subset of the first three glueball and meson states, we define the decoupled eigenenergies as 

\begin{equation}
\int \text{d}A \Psi_{nm}^T H_0 \Psi_{n'm'} = E_{nm} \delta_{nn'} \delta_{mm'} \, ,
\end{equation}
where $n,m=1,2,3$ and

\begin{equation*}
E_{nm} = \left( \begin{array}{c}
( E^{\zeta^{(0)}}_n)\\
( E^{\xi^{(0)}}_m)\\
\end{array} \right) \, ,
\end{equation*}
%
%
and $E_{nm}$ is given by the uncoupled energies, corresponding to meson masses of (1.688, 3.709, 4.975) and glueball masses of (3.846, 5.370, 6.386) in units of $\Lambda_{UV}$. The eigenfunctions of these first three meson and glueball states are shown in Fig. \ref{spect1}.

The total Hamiltonian in the basis of $\Psi_{nm}$ takes then the (infinite) matrix form, of which the part related to first 3 states of each kind is given by the following 6$\times$6  matrix

 {\footnotesize
 \begin{eqnarray}
&H_0+V =& \nn \\ 
&\left(\begin{tabular}{llllll}
17.6375  & 0.                  & 0.                  & -0.8702 & 0.2779   & -0.1221 \\
0.                  & 34.3941& 0.                  & -0.4534 & -1.4549  & 0.4314 \\
0.                  & 0.                  & 48.6373   & 0.2952 & -0.1115& -2.4406 \\
-9.2289  & -2.5654  & 2.5144  & 3.3982  & 0.                   & 0.                  \\
4.4342   & -12.3857& -1.4334 & 0.                  & 16.4069  & 0.                  \\
-0.5529& 1.0396   & -8.9644 & 0.                  & 0.                   & 29.5146
\end{tabular}\right) \, ,&
\end{eqnarray}}in which the diagonal values are the unmixed mass squared. The eigenvalues of this matrix give the new (mixed) energy values, from lowest to highest, which are:
\begin{equation}
E_{new}= (2.803, 15.056, 18.566, 28.380, 35.442, 49.741) \, ,
\end{equation}
which yields
\begin{equation}
m_{new}= (1.533, 3.553, 3.945, 4.878, 5.451, 6.458) \, ,
\end{equation}
in units of $\Lambda_{UV}$, where the first, second and fourth values correspond to the three lowest meson states, and the third, fifth, and sixth correspond to the glueballs (similar to what is shown in Fig. \ref{scalarmasses}). The coupled equations, on the other hand, gave the following masses for the first five mixed states: 1.533, 3.541, 3.943, 4.863, 5.447 in units of $\Lambda_{UV}$. 

\begin{figure}[!tb]
\begin{center}
\subfigure[]{
\includegraphics[width=0.49\textwidth]{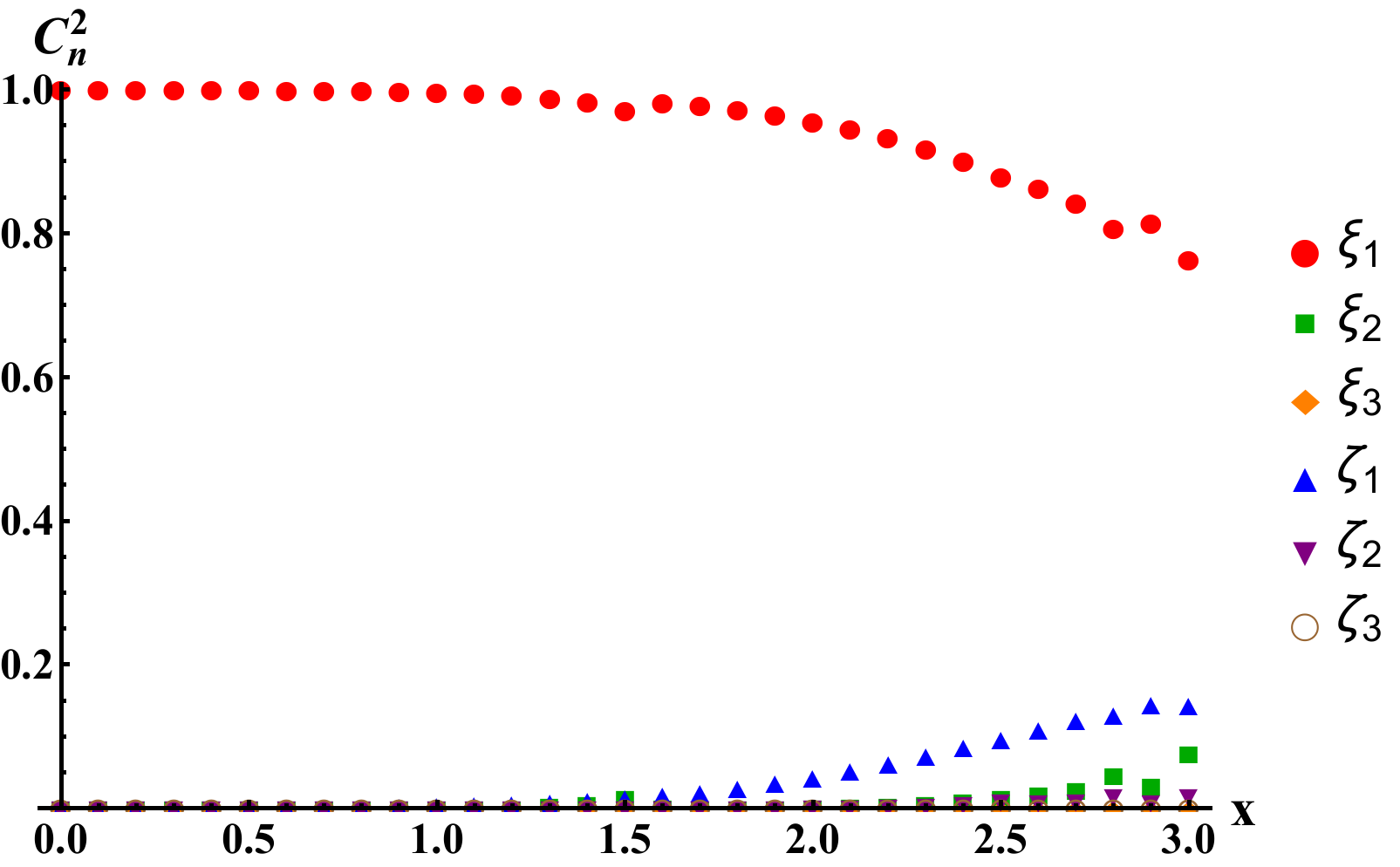}}
\subfigure[]{
\includegraphics[width=0.49\textwidth]{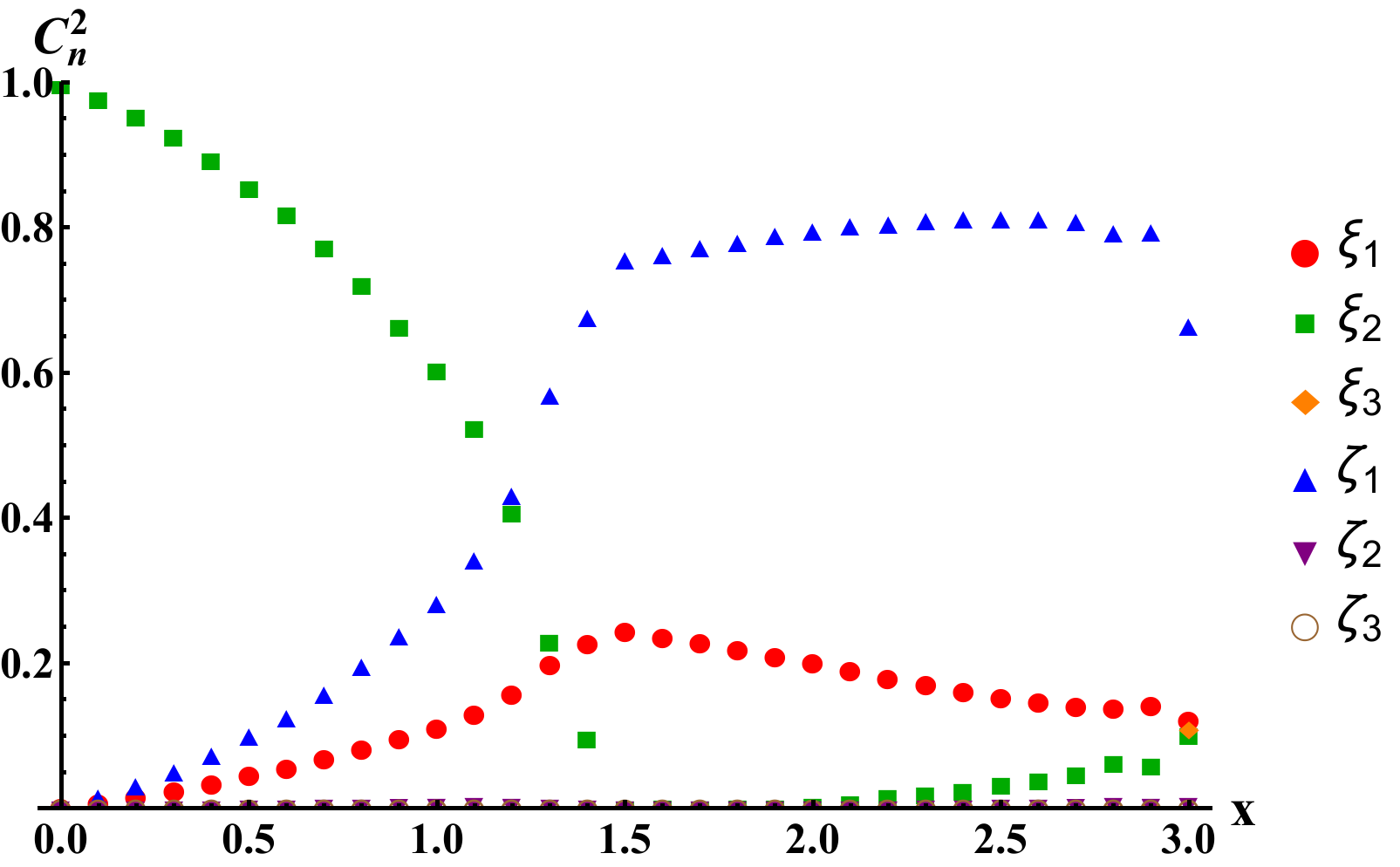}}
\end{center}
\caption{The square of the decomposition coefficients of the (a) lowest mixed meson state and (b) the lowest mixed glueball state.}
\label{mixfig}
\end{figure}

The eigenvectors can of course also be readily found. In principle those can be used to predict specific decay modes of the physical states. We do not do this, but only draw attention to the lowest of these states, the mixed sigma. Its decomposition, for $x=1$,  over these first six unmixed states reads

\begin{align}
 \ket{\sigma}&= .0587 \ket{1 0} +.0140 \ket{2 0} - 0.0065 \ket{3 0}+   0.9932 \ket{0 1} \nn \\
 & -0.0075 \ket{0 2} -0.0015 \ket{0 3}  \, ,
\end{align}
shown in Fig. \ref{mixfig} (a).
One finds that the unmixed lowest meson state $\ket{0 1}$ is very close to $\ket{\sigma}$, while the largest admixture to it is from the first glueball state $\ket{1 0}$. This statement however is not universal; while the second state vector is 
\begin{align}
 \ket{\sigma'}&= 0.0888 \ket{1 0} -0.0760 \ket{2 0}  -0.0022 \ket{3 0}-0.0541 \ket{0 1} \nn \\
 & -0.9916 \ket{0 2} +0.0074 \ket{0 3}  \, ,
\end{align}
which is close to the second unmixed meson, only
half of the third state is the unmixed first glueball
 \begin{align}
 \ket{f_0(1500)}& =0.5428 \ket{1 0} + 0.0599 \ket{2 0}  + 0.0085 \ket{3 0} \nn \\
 & -0.339 \ket{0 1}+0.7655 \ket{0 2} +0.0287 \ket{0 3}  \, ,
 \end{align}
 as observed in Fig. \ref{mixfig} (b) for $x=1$.

Note also that we have ignored many more states which can admix to these scalars, such as meson-glueball and  two-meson states; accounting for them will  reduce the ``original" fraction of the glueball a bit more.

The $x$-dependence of the mixing of the lowest mixed meson state and lowest mixed glueball state are shown in Figs. \ref{mixfig} (a) and (b), respectively. The plot depicts the squares of the mixing coefficients (found by taking the inner product of the mixed state with each unmixed eigenfunction and squaring). Particularly, Fig. \ref{mixfig} (b) explicitly shows that there is level-crossing at $x$ larger than one, as was first noticed in \cite{Arean:2012mq}.

The admixture of the glueball states to sigma is the key phenomenon we are interested in this paper. Indeed,
the strings -- as gluonic (dilatonic) objects -- do not directly interact with the quark-related brane (the $s$ scalar, in notations of  \cite{Arean:2013tja}).  One can express this fact with the string coupling vector $\ket{\text{string}}=(1,1,1,0,0,0)$ in the unmixed set of states. Its projection onto the physical sigma state is then
\be     \left<\sigma | \text{string}\right>  \approx 0.066 \ee    
which provides the key small parameter of the problem. This was first extracted in \cite{Kalaydzhyan:2014zqa} from lattice data of the two-string configuration.

%

\subsection{Scalar Fields From a String Source}

The Nambu-Goto action determines the string interaction with the metric and the dilaton. The excitations are rewritten here 
\begin{align}
\delta g_{MN}=\hat{g}_{MN} \,\, , \,\,\, \delta \Phi=\chi \, .
\end{align}
The first-order correction of the Nambu-Goto action due to the above excitation fields is

\begin{align}
S_{NG}&= -T_f \int dt dx \, e^{{4 \over 3} \Phi_0+ 2 A} \left( {4\over 3 }\chi+\right. \nn \\
& \left.  (\tilde G^{-1})^{\alpha \beta} \partial_\alpha X^{M} \partial_\beta X^{N} \hat{g}_{MN} \right)\, ,
\end{align}
where we have defined  $\tilde G_{\alpha \beta}= g_{\mu\nu} \partial_\alpha X^{\mu} \partial_\beta X^\nu$. The string is taken to be a $X^{M}=(t,x,X_2(t), X_3(t),Z(t))$. 
 In the case of low string velocities, only the dilaton is sourced, and the first-order action is simply

\be
S_{NG}= -T_f \int dt dx\, e^{2 A_s(z)} {4 \over 3} \chi(z) \, .
\label{NGsou}
\ee
Hence, at low velocities the NG Lagrangian, which sources the dilaton fluctuations is time independent.
The source term in the $\zeta$ equation of motion, Eq. (\ref{zeteq}), is given by the $\chi$ variation of the action, Eq. (\ref{NGsou})
\be
-{T_f \over 2 }e^{{4 \over 3} \Phi_0 -A} {A' \over \Phi_0'} \delta(z-z') \delta^2(r) \, ,
\ee
where $r=\sqrt{X_2^2+X_3^2}$. We now need to solve for the equations of motion sourced by the above delta-function-like perturbation. 

\be
 (H_0+V) \zeta(A,A') = \left( \begin{array}{c}
-w_\zeta(A) f_{NG}(A) \delta(A-A') \,, \\
0 \\
\end{array} \right)  \,, 
\ee
where $f_{NG}(A)={T_f \over 2} {e^{{4\over 3} \Phi(A)} e^{-A} \over \Phi'(A) z'(A') }$. Primes on the functions denote the derivative with respect to $A$, and $A'$ is the point where the source is located. This can be put in the matrix form described above, in which differential operators and delta functions are projected onto the set of unmixed states. We then have

\begin{align}
& \left( 
\begin{array}{cc}
H^{(0)}_\zeta +p^2 w_\zeta & V_\zeta \\
V_\xi & H^{(0)}_\xi +p^2 w_\xi
\end{array}
\right) 
\left(
\ba{c}
\zeta \\
\xi
\ea\right)   \nonumber   \\
& =\left( \begin{array}{c}
-w_\zeta(A) f_{NG}(A) \delta(A-A') \\
0 \\
\end{array} \right)  \, .
\label{gfe}
\end{align}
We expand the solutions and the source in terms of the $H^{(0)}$ eigenfunctions, 

\begin{equation*}
 \left(
\ba{c}
\zeta \\
\xi
\ea\right)=\sum c_n \chi_n=\sum \left(
\ba{c}
c^{(\zeta)}_n \zeta_n \\
 c^{(\xi)}_n \xi_n
\ea\right)
 \end{equation*} 
and 

\begin{equation*} 
f_{NG}(A) \delta(A-A')= \sum \dzn \zeta_n \, .
\end{equation*}
Then, Eq. (\ref{gfe}) becomes

\begin{align}
& \cz_\ell (m_\ell^{(\zeta)\, 2} +p^2) + \cxn  \int \zeta_\ell V_\zeta \xi_n = -\dz_\ell   \, ,\\
& \cx_\ell= -{\czn \int \xi_\ell V_\xi \zeta_n \over m_\ell^{(\xi)\, 2} +p^2}= -\czn {\bra{\xi_\ell}V_\xi \ket{\zeta_n} \over m_\ell^{(\xi)\, 2} +p^2}\, ,
\end{align}
We define $\int \zeta_\ell V_{\zeta/\xi} \xi_n=  \bra{\zeta_\ell}V_{\zeta/\xi} \ket{\xi_n}$ leading to

\begin{align}
\cz_\ell &=-\left( \delta_{\ell k}  +  {  \bra{\zeta_\ell}V_\zeta \ket{\xi_n} \,  \bra{\xi_n}V_\xi \ket{\zeta_k} \over \left(m_\ell^{(\zeta)\, 2} +p^2 \right) \left(m_n^{(\xi)\, 2} +p^2\right)}  \right) {\dz_k \over m_k^{(\zeta)\, 2} +p^2} \, ,\nn \\
\cx_\ell & = -{\dz_n \over m_n^{(\zeta)\, 2} +p^2} {\bra{\xi_\ell}V_\xi \ket{\zeta_n} \over m_\ell^{(\xi)\, 2} +p^2} \, ,
\end{align}
where $\dz_n=w_\zeta(A') \zeta_n(A') f_{NG}(A')$. The solutions then read

\begin{align}
\zeta(A,A')&=-{f_{NG}(A') w_\zeta(A') \zeta_n(A')  \zeta_k(A) \over m_n^{(\zeta)\, 2} +p^2} \times \nn \\ 
 &\left( \delta_{n k}  +  {  \bra{\zeta_n}V_\zeta \ket{\xi_\ell} \,  \bra{\xi_\ell}V_\xi \ket{\zeta_k} \over \left(m_\ell^{(\zeta)\, 2} +p^2 \right) \left(m_\ell^{(\xi)\, 2} +p^2\right)}  \right) \,,
\end{align}

\begin{align}
\xi(A,A')&= -{f_{NG}(A') w_\zeta(A') \zeta_n(A')  \xi_\ell (A) \over m_n^{(\zeta)\, 2} +p^2} {\bra{\xi_\ell}V_\xi \ket{\zeta_n} \over m_\ell^{(\xi)\, 2} +p^2}  \, .
\end{align}


Yukawa potentials are given by the propagator, $\sim\exp(-m_n r)$. Since the distances of interest are $r\sim 1$ fm, going from the first to the second scalar means an extra factor of  $\sim \exp[-(m_2-m_1) r]\sim \exp[-5] \ll 1 $. Thus, one can safely ignore all scalar states except the first one.


The solution in coordinate space reads
\be 
\zeta(z,x,z',x') = \sum_n \int {d^4p \over (2\pi)^4 }  e^{ip_i (x-x')^i} \zeta \left(A(z),A(z'),p \right) \,.
\ee 
Considering strings in the quasi-static limit, $p_0=0$, and parallel in $x^1$ direction with $p_1=0$ , one is left with only two transverse momenta $p_2,p_3$ over which to integrate. In the limit of $p\ll m^{(\zeta)}_n$, the first-order $\zeta$ field term gives a contact term in the $x^2x^3$ plane, so we keep only the second-order term to describe long-range interaction. The  potential is found to be a Yukawa potential in the $x^2x^3$ plane of the form of the one used in  \cite{Kalaydzhyan:2014zqa} with additional dependence on the $z$-coordinate, which is governed by the holographic dynamics.  As shown above, the string couples to the $\chi$ field. Selecting the gauge where $\psi=0$, we have that $\chi(z)=-{A'(z) \over \Phi'(z)} \zeta(z) $.

\begin{align}
\chi(z,z',r)= -& \bra{\zeta_1}V_\zeta \ket{\xi_1} \,  \bra{\xi_1}V_\xi \ket{\zeta_1} \times  \nonumber    \\  
&{ f_{NG}(z')  w_\zeta(z')  \zeta_1(z') \zeta_1(z) \over 2 \pi \, m_1^{(\zeta)\, 4}} {A'(z) \over \Phi'(z)} K_0(m^{(\xi)}_1 \,r) \, .
\end{align}

We now consider string motion in the background potential and the $\chi$ fluctuation. The string action is 
\be
S_{NG}= -T_f \int dt dx\, e^{2 A_s(z)}\left(1+{4 \over 3} \chi\right)\sqrt{1- \dot Z^2 -\dot X_2^2-\dot X_3^2} \,.
\ee
In the non-relativistic limit, the equations of motion are

\begin{align}
\ddot Z(t)&=-2 \partial_Z A_s(Z)-{4 \over 3}\partial_Z \chi(Z,X_2,X_3) \,, \nonumber \\
\ddot X_2(t)&=-{4\over 3} \partial_{X_2} \chi(Z,X_2,X_3) \,, \nonumber  \\
 \ddot X_3(t)&=-{4\over 3} \partial_{X_3} \chi(Z,X_2,X_3) \, .
\end{align}

\section{Chiral Symmetry Restoration}
\label{chsymre}

 As was shown initially in  \cite{ckp} and further studied in \cite{ikp, jk}, chiral symmetry breaking is described by the dynamics of the bulk tachyon field, which is dual to the $\bar q q$ operator. The diverging IR asymptotics of the tachyon space-time lead to the breaking of chiral symmetry in the boundary field theory. The chiral condensate is read holographically by the UV asymptotics of the bulk tachyon field, Eq. (\ref{tachuv}).
 
The Nambu-Goto string, in the gravity-dilaton-tachyon background, sources the scalar glueball fluctuation ($\chi$), which is also coupled to $\sigma$-meson ($s$), Eqs. (\ref{zeteq}, \ref{xieq}). In the case of a large number of string sources, the excitations are strengthened and they can significantly change the background dilaton and tachyon fields. 

   \begin{figure}[t]
  \begin{center}
  \subfigure[]{
  \includegraphics[width=.49\textwidth]{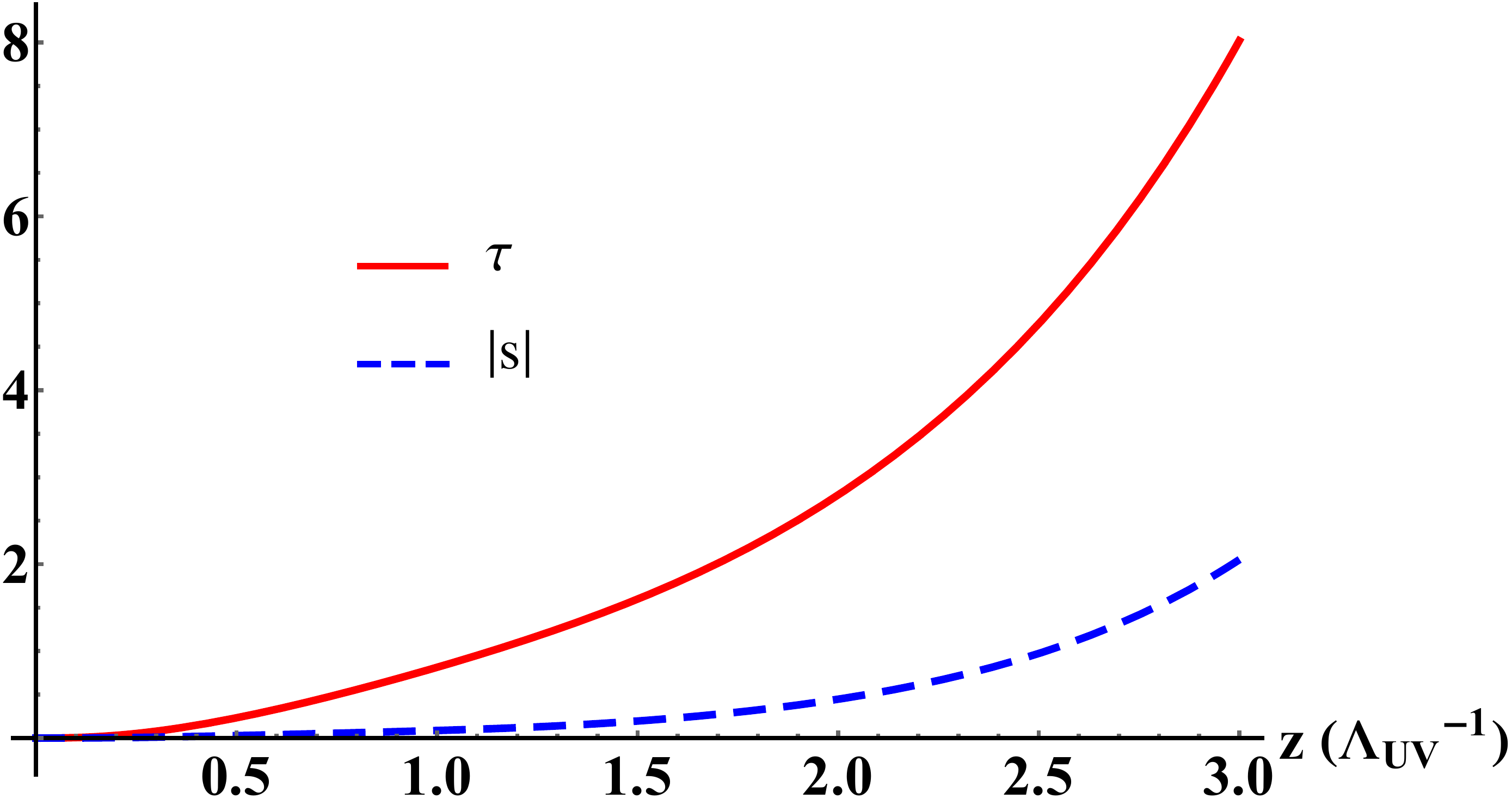}}
  \subfigure[]{
    \includegraphics[width=.49\textwidth]{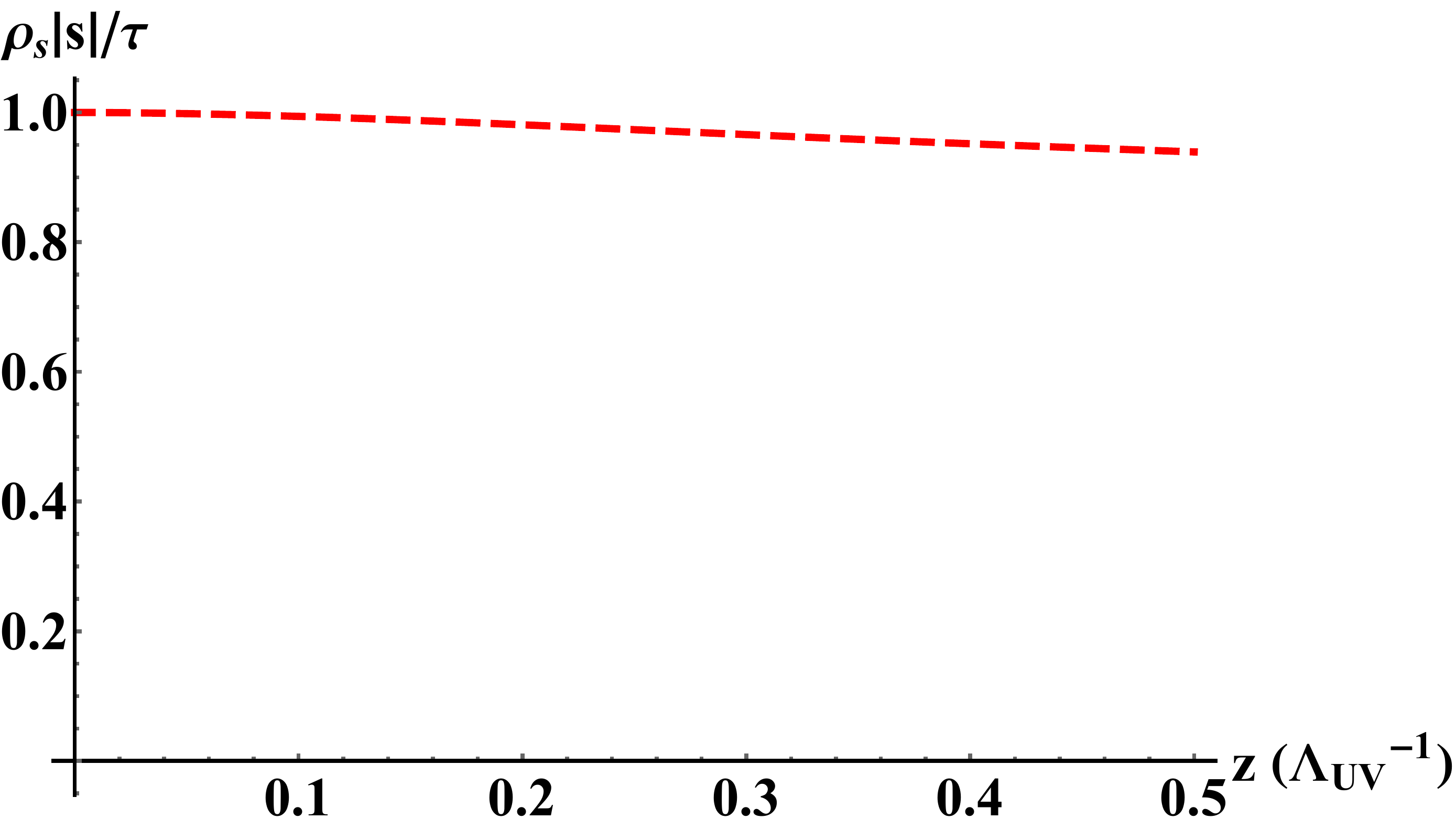}}
   \caption{(a) The (tachyon) excitation induced by a single string $s$ is compared to the vacuum (background) field.
   (b) The ratio of the (tachyon) excitation produced by transverse density of strings $\rho_s=10.85$ fm$^{-2}$ with the vacuum (background) field.
   }
  \label{eva}
  \end{center}
\end{figure}

The fluctuation induced by one string is compared to the vacuum solution  in Fig. \ref{eva}(a),
for a source string at the equilibrium point  $z_*$.
As one can see, the $z$-dependence of both is similar, but the effect of one string is small compared
to the background.

Placing static strings around the equilibrium point, $z_*$, we calculate the necessary density of strings that source an excitation field that cancels the vacuum solution close to the boundary. The density of strings is defined as the number of strings over the area that they occupy. We notice that since the string sources are static, the fluctuation field will also be time independent. The ratio of the two, for the critical string density $\rho_s=10.85$ fm$^{-2}$, is shown in  Fig. \ref{eva}(b); one can see that it is close to unity for a large range of $z$. This density of coincidental strings thus cancels the vacuum solution, and consequently leads to restoration of the chiral symmetry, similar to other models \cite{Kalaydzhyan:2014zqa, iritani}; this restoration is unlike that of those studies, however, as they observe the modification of the chiral condensate within the flux tube.

\begin{figure}[!tb]
  \begin{center}
  \includegraphics[width=.49\textwidth]{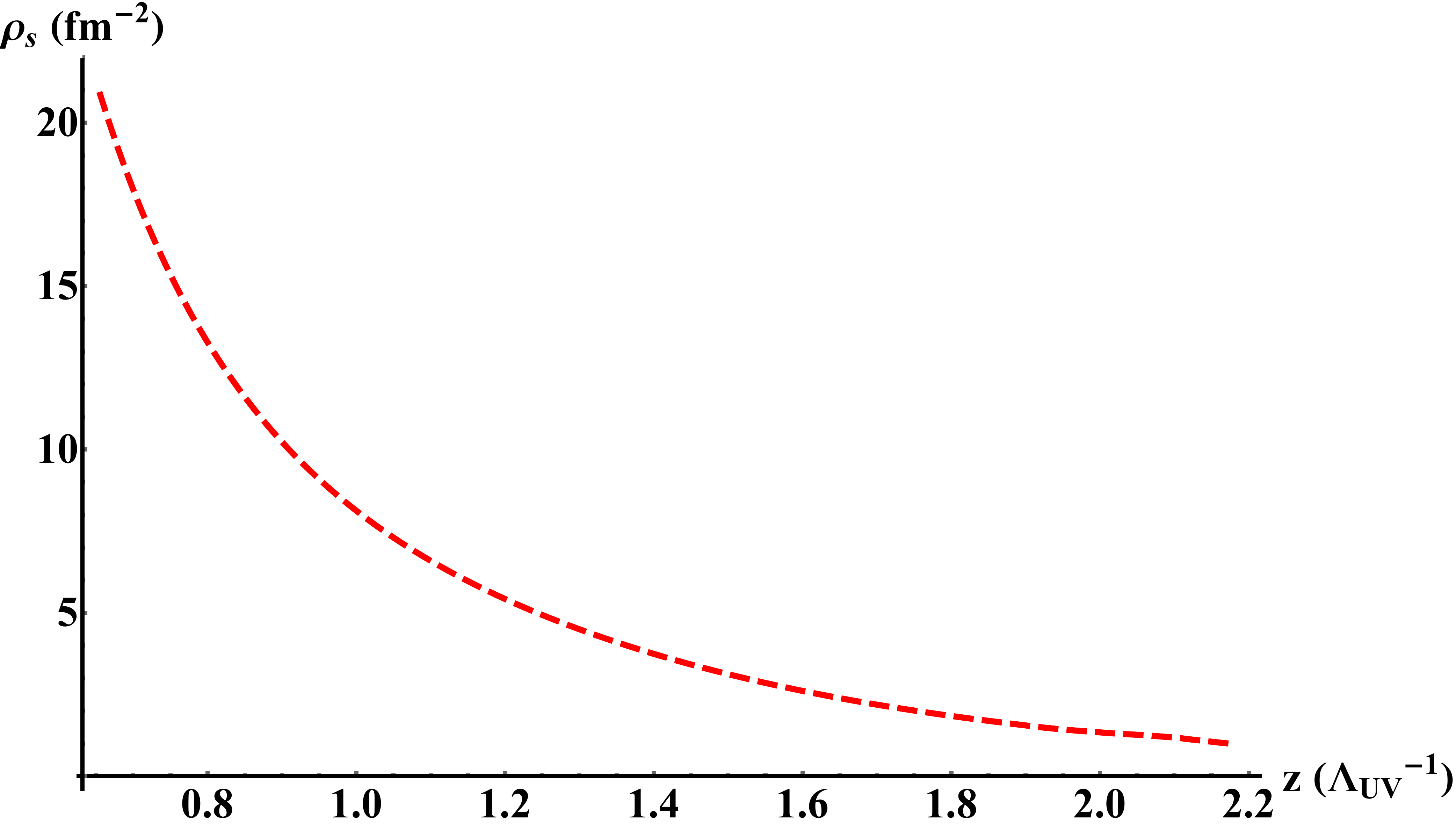}
    \end{center}
  \caption{The density of strings necessary to induce a (tachyon) excitation that cancels the vacuum (background) field as a function of placement in the $z$ coordinate.}
  \label{csz}
  \end{figure}

Since, in this study, we are concerned with the interaction and dynamics of the strings, the strings will not always be at the equilibrium point of the potential. Due to the nature of the $\chi$ fluctuation and its dependence on the source point $z'$, the density of strings that are needed to source an excitation that cancels the vacuum field varies with location. This behavior is shown in Fig. \ref{csz}. One can see that if the strings are located further into the infrared (larger $z$), their individual contribution is larger, and thus, a lesser density of strings is necessary to restore chiral symmetry.

The numbers given are based on a simple linear extrapolation from a single string. Obviously, when the  excitation field that gets comparable to the vacuum field, the problem becomes nonlinear; the sources should be included in the vacuum equations of motion. To solve those is beyond the scope of the present work.

\section{Collective effect of strings in the bulk background}

\begin{figure}[!tb]
\begin{center}
\subfigure[]{
\includegraphics[width=0.49\textwidth]{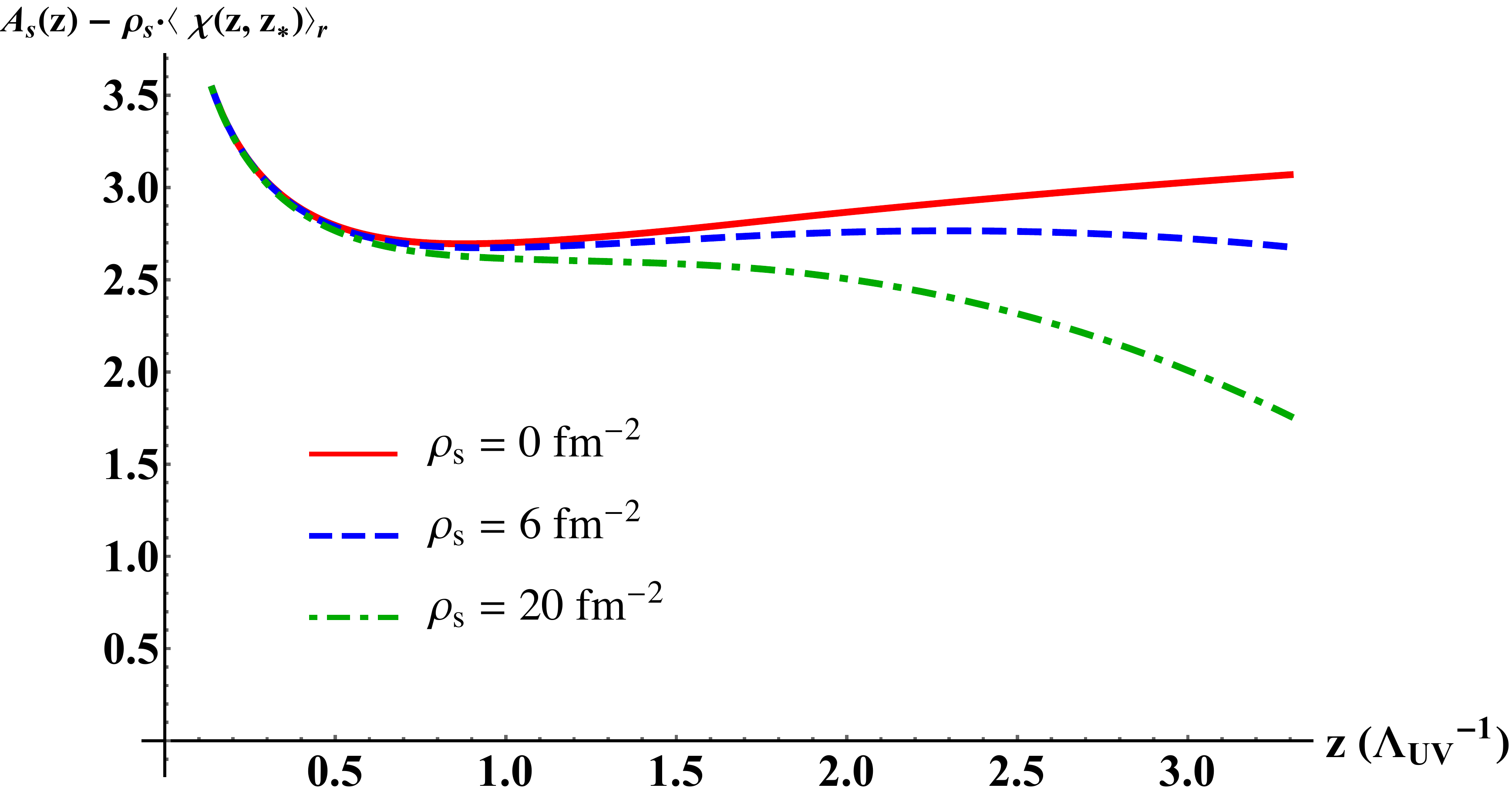}} \quad
\subfigure[]{\includegraphics[width=0.49\textwidth]{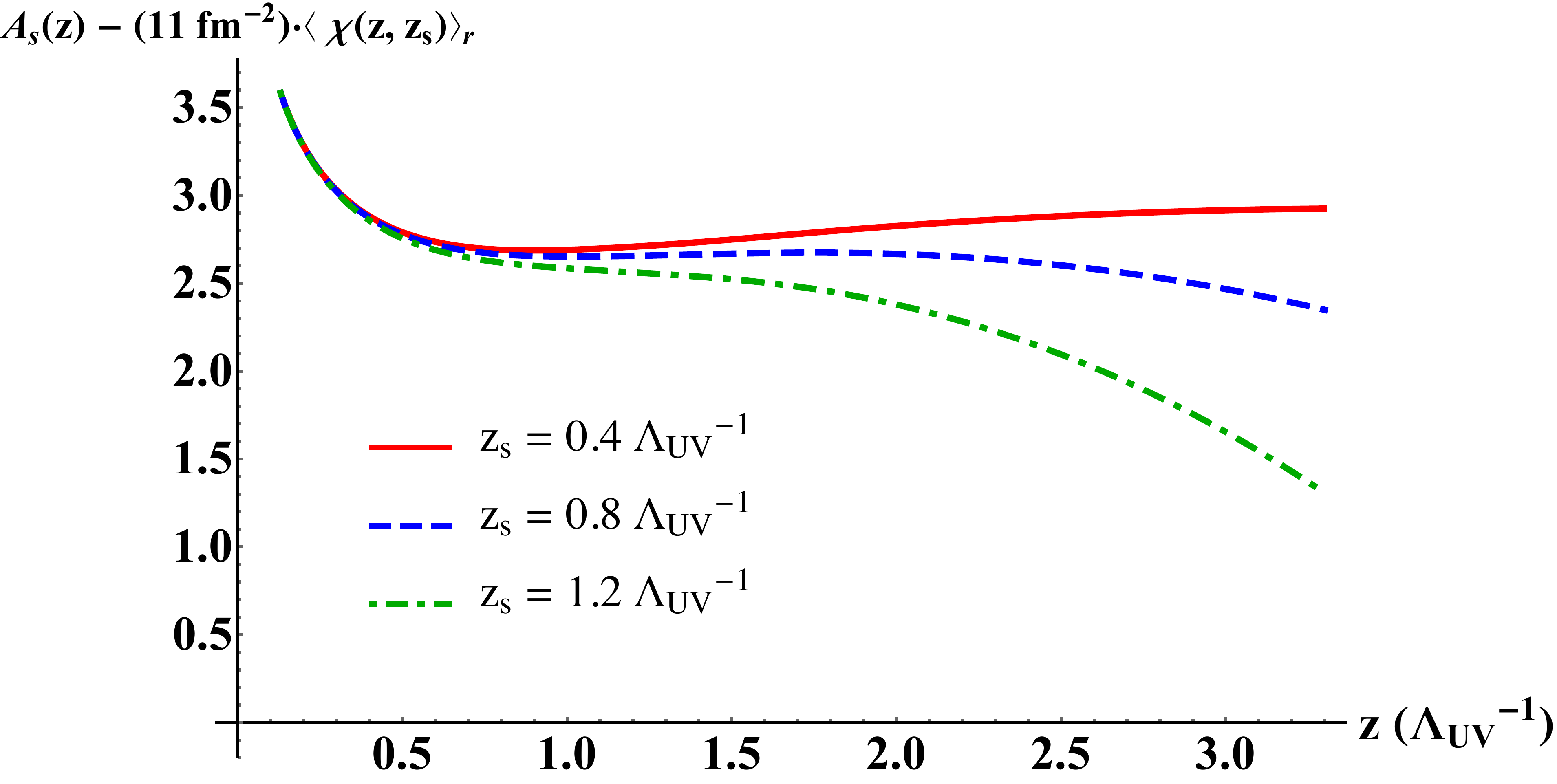}}
\end{center}
\caption{The background potential, (a) without and with string-induced fluctuations, all placed at the minimum of the $z$ potential ($z_*$) with the denoted transverse density, and (b) induced by strings with density 11 fm$^{-2}$, all placed at various points in the $z$ coordinate (denoted $z_s$). The $r$ dependence of $\chi$ is averaged out, leaving only the density dependence of the fluctuation.}
\label{bkg}
\end{figure}

In this section, we consider effects of strings, distributed in the transverse 2-d plane with a certain density, on the background potential. The string fluctuation deformation of the background due to different densities is shown in Fig. \ref{bkg}(a). We notice that the minimum of the potential no longer exists if a certain density of strings is present at $z = z_*$ (approximately 18 fm$^{-2}$).

Similarly, we also notice that, based on the form of $\chi(z, z')$, the further a string is placed from the holographic boundary, the larger the induced fluctuation on the background. This behavior is shown in Fig. \ref{bkg}(b), in which the string fluctuation is sourced by strings with transverse density 11 fm$^{-2}$. The strings are then placed at a certain point in the $z$ coordinate, denoted $z_s$. We see that at $z_s = 0.4\,\Lambda_{UV}^{-1}$, the fluctuation is almost negligible -- the background remains almost unchanged from the zero string case --, but if the strings are placed at $z_s = 1.2\,\Lambda_{UV}^{-1}$, the string induced string fluctuation is equivalent to a much larger density of strings placed at the original minimum of the potential; compare with Fig. \ref{bkg}(a).

In the original background with no string sources, a string would be in a stable oscillating state when initially placed anywhere in the holographic direction. However, with the induced fluctuations of a number of strings, the string may be in, or move to, an unstable point of the potential and subsequently be pushed indefinitely far from the UV boundary ($z=0$). The configurations that exhibit this behavior are highly dependent on initial conditions: the number of strings, initial placement in the $z$-direction, and the transverse density. As can be seen in the equations of motion, the strings attract in the transverse plane, while they ``fall" in the holographic coordinate. Therefore, when enough strings are close enough and induce a large enough fluctuation, on the order of magnitude of the original background potential, they collapse onto each other in the transverse plane and are repelled infinitely far from the UV boundary in the $z$-direction (i.e. ``fall'' into the IR). When this collapse occurs, the interaction of strings is no longer described by our model, as they collectivize into a black hole and back-reaction must be taken into account.

\section{Numerical Study of Multi-String Motion}

To study the motion and interaction of a number of strings, we wrote a program in Python, utilizing its odeint differential equation module. The background functions, wavefunctions, and subsequent interaction functions were computed and analyzed in Mathematica and exported for use in the Python program.

The initial conditions in the transverse ($x^2$, $x^3$) coordinates are the same as those given in \cite{Kalaydzhyan:2014zqa}. In the $z$-direction, a simple Gaussian distribution was used with a standard deviation of 0.11 $\Lambda_{UV}^{-1}$. It is important to note that, in the $z$-direction, the interaction does not have to do with the actual distance between a fluctuation-sourcing string and any other string; the position of a sourcing string determines it's effect on the background potential, which then changes the other strings' behavior.

Using these initial conditions, the number of ``wounded" nucleons, $N_w$, each of which creates two strings, was varied. The cases treated were $N_w = 2,3,4,5,6,7,8,9,10,11,13,15,18,20, \text{ and } 25$, each of which had the transverse distribution described above and centered at $z = 0.44, 0.55, 0.66, 0.77, 0.88, 0.99, 1.1, \text{ and } 1.2$ $\Lambda_{UV}^{-1}$. Sixteen trials were run for each string number and initial $z$ distribution.

\begin{figure}[!tb]
\begin{center}
\includegraphics[width=0.49\textwidth]{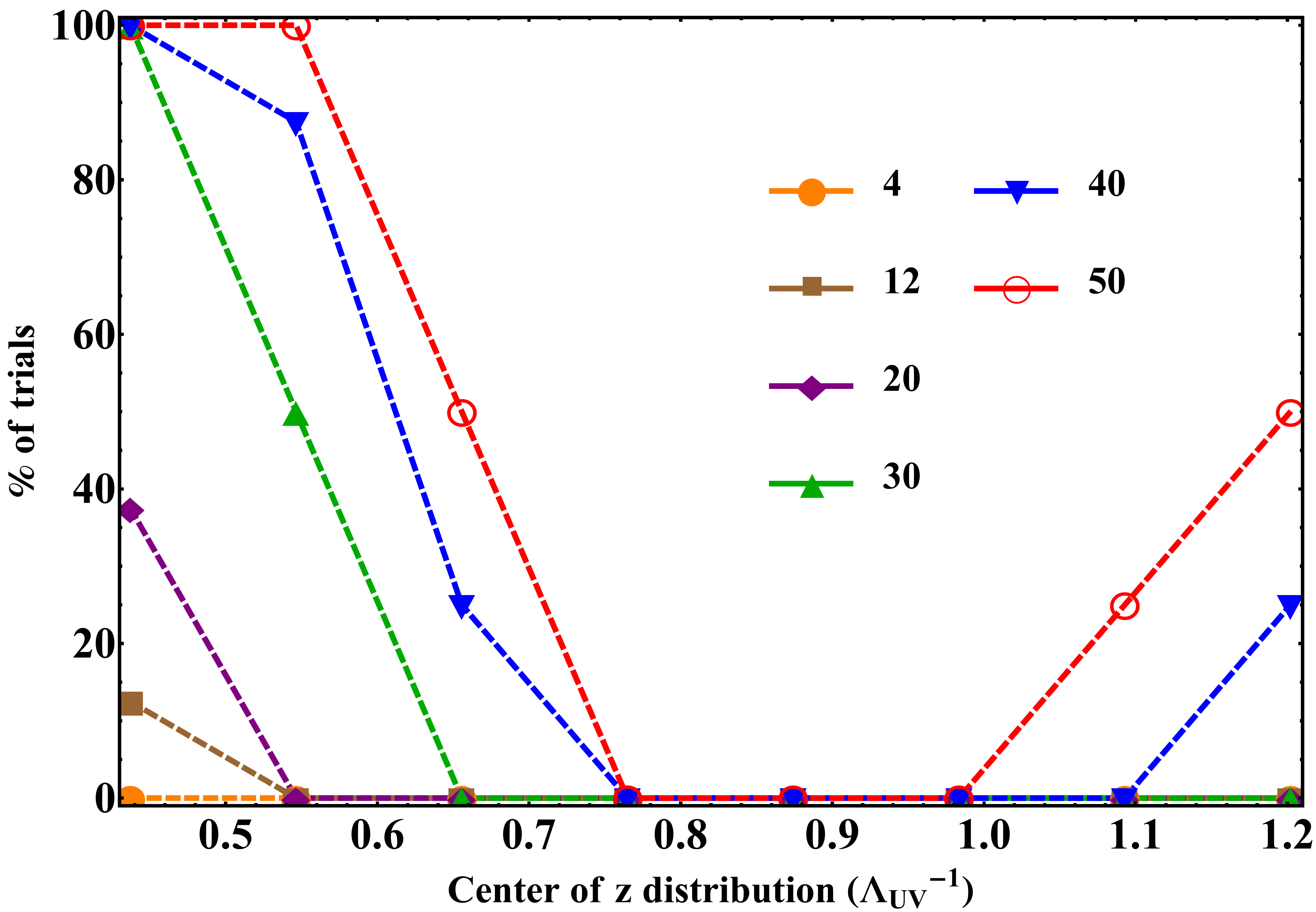}
\end{center}
\caption{Percentage of trials in which the strings collectivized before breaking (i.e. within 1.5 fm/c) as a function of number of strings and initial center of Gaussian distribution in $z$.}
\label{pcts}
\end{figure}

\begin{figure*}
\begin{center}
\subfigure[]{%
\includegraphics[width=0.78\textwidth]{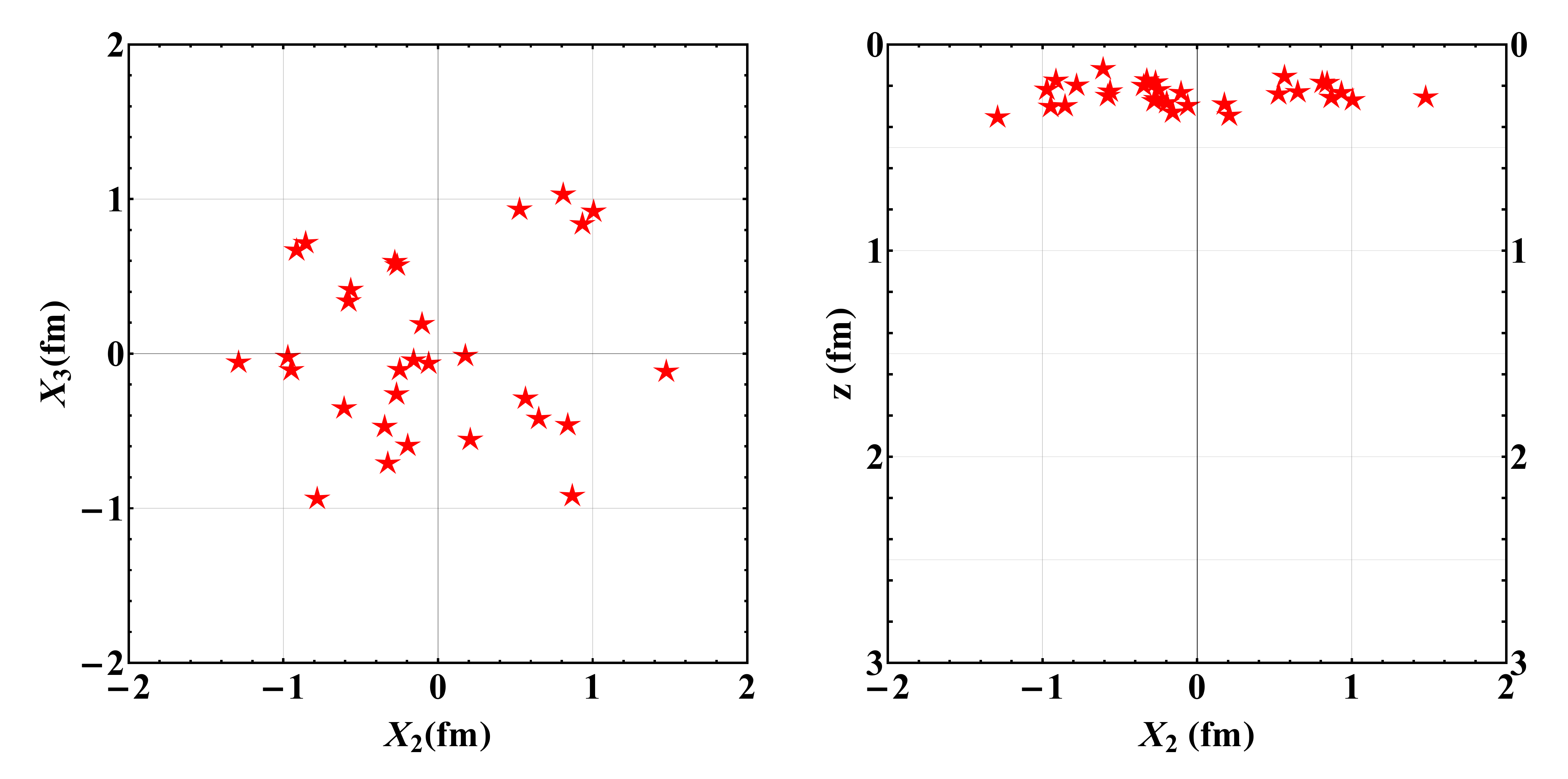}} \quad
\subfigure[]{
\includegraphics[width=0.78\textwidth]{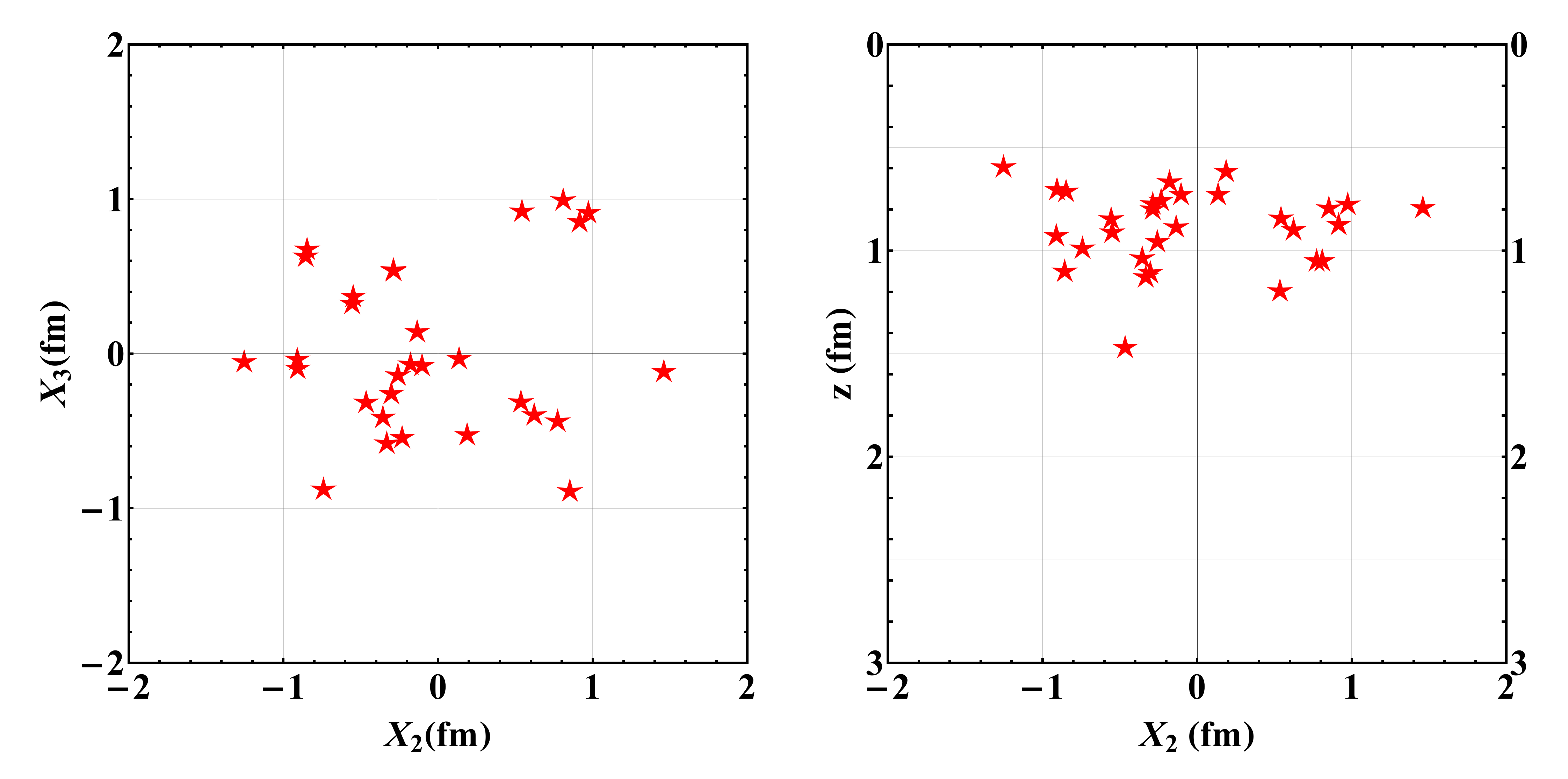}} \quad
\subfigure[]{
\includegraphics[width=0.78\textwidth]{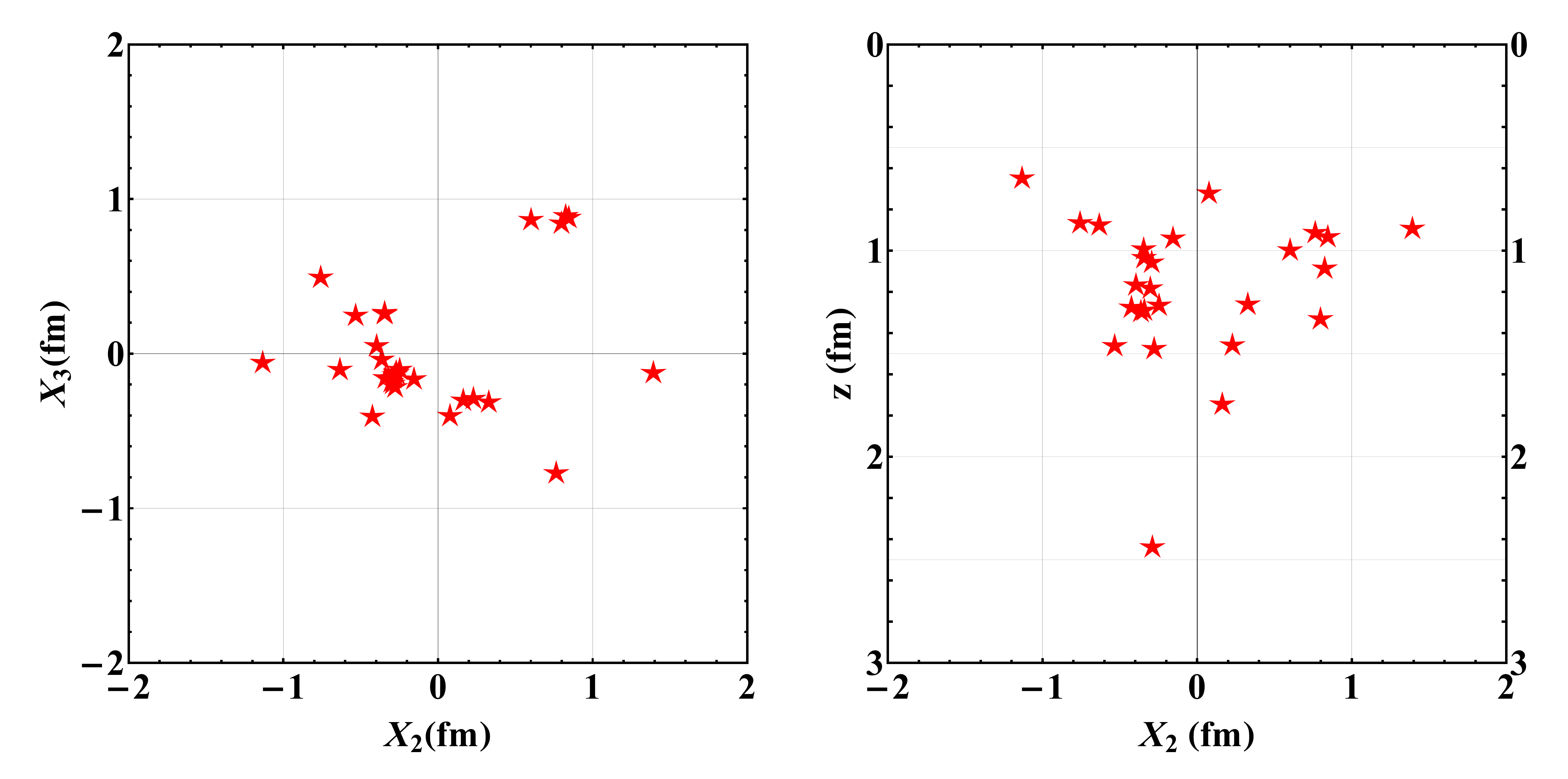}} \quad

\end{center}
\caption{Snapshots in the transverse (left) and holographic (right) planes of a falling 30 string configuration initially centered at $z= 0.44\,\Lambda_{UV}^{-1}$ at $t=$ (a) 0 fm/c, (b) 1.0 fm/c, and (c) 1.32 fm/c.}
\label{exs}
\end{figure*}

The timescale of interest is on the order of 1.5 fm/c. By the end of this time period, the strings break due to Schwinger type $\bar{q}q$ production and subsequently disappear, as was mentioned above.

The collectivization results are shown in Fig. \ref{pcts}. We see that the higher momentum states -- those that are initially closer to the UV boundary -- have a higher probability to collectivize with a fewer number of strings in the time period of interest. Cases with $N_w =$ 20-25 collectivize regularly when initially close to the boundary but do not ``fall" before breaking if they are near the minimum of the original potential. 

However, we also see that if they are far enough away from the boundary (around $z=1.1\,\Lambda_{UV}^{-1}$ for 50 strings, for example), they once again begin to fall before breaking due to their larger induced fluctuation on the background potential (see Fig. \ref{bkg}(b)).

An example of a configuration of strings that falls is shown in Fig. \ref{exs}. This configuration of 30 strings centered around 0.44 $\Lambda_{UV}^{-1}$ collectivizes in approximately 1.4 fm/c, which is when the strings all ``fall" to the infrared (large $z$) region. We cannot see any behavior after 1.32 fm/c in this particular trial because at least one of the strings falls beyond the IR cutoff of our numerics. The perturbation on the background caused by this string located at large $z$ is large, since large $z$ magnifies the effect of each string on the background, as shown in Fig. \ref{bkg}(b). In this case, the background potential no longer has a minimum, so we see that as soon as one of the strings is far enough from the $z=0$ boundary, the induced background fluctuation causes all other strings to subsequently follow the first to the IR.

\section{Summary} 
The aim of this work was to study string dynamics in a holographic setting. We have chosen to use the V-QCD holographic model, with background potentials of class I given in \cite{Arean:2013tja}. It is worth noting that we have not changed any parameters or potentials of the model in this work. In this sense, all we do is a straightforward calculation, in a framework fully defined by matching to other QCD features in previous works \cite{jk, Alho:2012mh, Arean:2012mq, Arean:2013tja, Iatrakis:2014txa, Jarvinen:2015ofa}. 

We studied the spectroscopy of the scalar hadrons  by treating the full states as a mixture of uncoupled meson and glueball states, which are coupled perturbatively.  We found that the perturbative spectra are in a particularly good agreement with the full numerical solution of the V-QCD model. We also found interesting level-crossing between glueballs and mesons.

We then turned to QCD strings, the holograms of the fundamental strings in the bulk, levitating in the
minimum of the effective potential formed by gravitational and dilaton fields. The first qualitative result we obtain is that the QCD strings should have a ``gluonic core," associated with glueball mass, and a ``sigma cloud," associated with the sigma meson mass.

We then found the response of all fields due to the presence of a string by considering the coupling between the scalar glueballs and mesons to be perturbative. This approach is  supported by the fact that the perturbative spectrum is very close to that of the fully coupled system. We found that a certain density of strings cancels the chiral condensate in the vacuum background, thus restoring the chiral symmetry. The magnitude of string-string interaction is related to the strength of the glueball-meson mixing. This is weak because, in the no-mixing approximation, strings do not interact with quark-made mesons. 

We also found that corrections to the dilaton background can modify it in such a way that the minimum of the effective string potential disappears. If so, there are no more QCD strings, since all fundamental strings are  indefinitely falling into the IR direction.   

Finally, we considered motion of the multi-string system, initiated in a ``spaghetti" configuration, in which they are all considered to be extended indefinitely in the ``beam" direction, so that the motion is only happening in 3 transverse coordinates: the plane transverse to the beam and the holographic coordinate $z$. Depending on the number of strings and their initial locations, one observes attraction between them, which can be strong enough to increase the local density by an order of magnitude or so, and therefore trigger the collective ``falling" into the IR direction. This is physically interpreted as ``string melting" into a QGP fireball, as is indeed observed in $pA$ collisions of high enough multiplicity. 

\vskip 1cm
{\bf Acknowledgements.}
We would like to thank E. Kiritsis for useful discussions. This work was supported in part by the U.S. Department of Energy under Contract No. DE-FG-88ER40388.

\end{document}